\documentclass[pra,aps,showpacs,onecolumn,twoside,superscriptaddress]{revtex4-1}

\usepackage{amsmath,amsfonts,amssymb,caption,color,epsfig,graphics,graphicx,hyperref,latexsym,mathrsfs,revsymb,theorem,url,verbatim,epstopdf}

\usepackage{amssymb}
\usepackage{color}
\usepackage{amsmath,epic,curves,amscd}
\usepackage[english]{babel}
\usepackage{graphicx}
\usepackage{comment}
\usepackage{appendix}
\usepackage{mathdots}

\newtheorem{definition}{Definition}
\newtheorem{proposition}[definition]{Proposition}
\newtheorem{lemma}[definition]{Lemma}

\newtheorem{theorem}[definition]{Theorem}
\newtheorem{corollary}[definition]{Corollary}
\newtheorem{conjecture}[definition]{Conjecture}

\newtheorem{remark}[definition]{Remark}
\newtheorem{example}[definition]{Example}
\newtheorem{question}[definition]{Question}

\def\squareforqed{\hbox{\rlap{$\sqcap$}$\sqcup$}}
\def\qed{\ifmmode\squareforqed\else{\unskip\nobreak\hfil
\penalty50\hskip1em\null\nobreak\hfil\squareforqed
\parfillskip=0pt\finalhyphendemerits=0\endgraf}\fi}
\def\endenv{\ifmmode\;\else{\unskip\nobreak\hfil
\penalty50\hskip1em\null\nobreak\hfil\;
\parfillskip=0pt\finalhyphendemerits=0\endgraf}\fi}
\newenvironment{proof}{\noindent \textbf{{Proof.~} }}{\qed}
\def\Dbar{\leavevmode\lower.6ex\hbox to 0pt
{\hskip-.23ex\accent"16\hss}D}
\makeatletter
\def\url@leostyle{%
  \@ifundefined{selectfont}{\def\UrlFont{\sf}}{\def\UrlFont{\small\ttfamily}}}
\makeatother
\def\bcj{\begin{conjecture}}
\def\ecj{\end{conjecture}}
\def\bcr{\begin{corollary}}
\def\ecr{\end{corollary}}
\def\bd{\begin{definition}}
\def\ed{\end{definition}}
\def\bea{\begin{eqnarray}}
\def\eea{\end{eqnarray}}
\def\bem{\begin{enumerate}}
\def\eem{\end{enumerate}}
\def\bex{\begin{example}}
\def\eex{\end{example}}
\def\bim{\begin{itemize}}
\def\eim{\end{itemize}}
\def\bl{\begin{lemma}}
\def\el{\end{lemma}}
\def\bma{\begin{bmatrix}}
\def\ema{\end{bmatrix}}
\def\bpf{\begin{proof}}
\def\epf{\end{proof}}
\def\bpp{\begin{proposition}}
\def\epp{\end{proposition}}
\def\bqu{\begin{question}}
\def\equ{\end{question}}
\def\br{\begin{remark}}
\def\er{\end{remark}}
\def\bt{\begin{theorem}}
\def\et{\end{theorem}}

\def\btb{\begin{tabular}}
\def\etb{\end{tabular}}

\newcommand{\nc}{\newcommand}

\def\a{\alpha}
\def\b{\beta}

\def\m{\mu}
\def\n{\nu}
\def\x{\xi}

\def\s{\sigma}

\def\T{\Theta}

\def\S{\Sigma}

 \nc{\bbA}{\mathbb{A}} \nc{\bbB}{\mathbb{B}} \nc{\bbC}{\mathbb{C}}
 \nc{\bbD}{\mathbb{D}} \nc{\bbE}{\mathbb{E}} \nc{\bbF}{\mathbb{F}}
 \nc{\bbG}{\mathbb{G}} \nc{\bbH}{\mathbb{H}} \nc{\bbI}{\mathbb{I}}
 \nc{\bbJ}{\mathbb{J}} \nc{\bbK}{\mathbb{K}} \nc{\bbL}{\mathbb{L}}
 \nc{\bbM}{\mathbb{M}} \nc{\bbN}{\mathbb{N}} \nc{\bbO}{\mathbb{O}}
 \nc{\bbP}{\mathbb{P}} \nc{\bbQ}{\mathbb{Q}} \nc{\bbR}{\mathbb{R}}
 \nc{\bbS}{\mathbb{S}} \nc{\bbT}{\mathbb{T}} \nc{\bbU}{\mathbb{U}}
 \nc{\bbV}{\mathbb{V}} \nc{\bbW}{\mathbb{W}} \nc{\bbX}{\mathbb{X}}
 \nc{\bbZ}{\mathbb{Z}}

 \nc{\bA}{{\bf A}} \nc{\bB}{{\bf B}} \nc{\bC}{{\bf C}}
 \nc{\bD}{{\bf D}} \nc{\bE}{{\bf E}} \nc{\bF}{{\bf F}}
 \nc{\bG}{{\bf G}} \nc{\bH}{{\bf H}} \nc{\bI}{{\bf I}}
 \nc{\bJ}{{\bf J}} \nc{\bK}{{\bf K}} \nc{\bL}{{\bf L}}
 \nc{\bM}{{\bf M}} \nc{\bN}{{\bf N}} \nc{\bO}{{\bf O}}
 \nc{\bP}{{\bf P}} \nc{\bQ}{{\bf Q}} \nc{\bR}{{\bf R}}
 \nc{\bS}{{\bf S}} \nc{\bT}{{\bf T}} \nc{\bU}{{\bf U}}
 \nc{\bV}{{\bf V}} \nc{\bW}{{\bf W}} \nc{\bX}{{\bf X}}
 \nc{\bZ}{{\bf Z}}

\nc{\cA}{{\cal A}} \nc{\cB}{{\cal B}} \nc{\cC}{{\cal C}}
\nc{\cD}{{\cal D}} \nc{\cE}{{\cal E}} \nc{\cF}{{\cal F}}
\nc{\cG}{{\cal G}} \nc{\cH}{{\cal H}} \nc{\cI}{{\cal I}}
\nc{\cJ}{{\cal J}} \nc{\cK}{{\cal K}} \nc{\cL}{{\cal L}}
\nc{\cM}{{\cal M}} \nc{\cN}{{\cal N}} \nc{\cO}{{\cal O}}
\nc{\cP}{{\cal P}} \nc{\cQ}{{\cal Q}} \nc{\cR}{{\cal R}}
\nc{\cS}{{\cal S}} \nc{\cT}{{\cal T}} \nc{\cU}{{\cal U}}
\nc{\cV}{{\cal V}} \nc{\cW}{{\cal W}} \nc{\cX}{{\cal X}}
\nc{\cY}{{\cal Y}}
\nc{\cZ}{{\cal Z}}

\nc{\hA}{{\hat{A}}} \nc{\hB}{{\hat{B}}} \nc{\hC}{{\hat{C}}}
\nc{\hD}{{\hat{D}}} \nc{\hE}{{\hat{E}}} \nc{\hF}{{\hat{F}}}
\nc{\hG}{{\hat{G}}} \nc{\hH}{{\hat{H}}} \nc{\hI}{{\hat{I}}}
\nc{\hJ}{{\hat{J}}} \nc{\hK}{{\hat{K}}} \nc{\hL}{{\hat{L}}}
\nc{\hM}{{\hat{M}}} \nc{\hN}{{\hat{N}}} \nc{\hO}{{\hat{O}}}
\nc{\hP}{{\hat{P}}} \nc{\hR}{{\hat{R}}} \nc{\hS}{{\hat{S}}}
\nc{\hT}{{\hat{T}}} \nc{\hU}{{\hat{U}}} \nc{\hV}{{\hat{V}}}
\nc{\hW}{{\hat{W}}} \nc{\hX}{{\hat{X}}} \nc{\hZ}{{\hat{Z}}}

\nc{\hn}{{\hat{n}}}

\def\max{\mathop{\rm max}}
\def\min{\mathop{\rm min}}

\def\UOM{{\mbox{\rm UOM}}}

\def\lf{\lfloor}
\def\rf{\rfloor}

\def\ox{\otimes}

\newcommand{\ket}[1]{|#1\rangle}

\newcommand{\tbc}{\red{TO BE CONTINUED...}}

\newcommand{\opp}{\red{OPEN PROBLEMS}.~}

\newcommand{\red}{\textcolor{red}}

\newcommand{\jpa}{J. Phys. A}

\def\Dbar{\leavevmode\lower.6ex\hbox to 0pt
{\hskip-.23ex\accent"16\hss}D}

\begin{document}

\title{The unextendible product bases of four qubits: Hasse diagrams}

\author{Lin Chen}
\email{linchen@buaa.edu.cn (corresponding author)}
\affiliation{School of Mathematics and Systems Science, Beihang University, Beijing 100191, China}
\affiliation{International Research Institute for Multidisciplinary Science, Beihang University, Beijing 100191, China}

\def\Dbar{\leavevmode\lower.6ex\hbox to 0pt
{\hskip-.23ex\accent"16\hss}D}
\author {{ Dragomir {\v{Z} \Dbar}okovi{\'c}}}
\email{djokovic@uwaterloo.ca}
\affiliation{Department of Pure Mathematics and Institute for
Quantum Computing, University of Waterloo, Waterloo, Ontario, N2L 3G1, Canada} 

\date{\today}


\begin{abstract}
We consider the unextendible product bases (UPBs) of fixed cardinality $m$ in quantum systems of $n$ qubits. These UPBs are 
divided into finitely many equivalence classes with respect to 
an equivalence relation introduced by N. Johnston. There is a natural partial order ``$\le$'' on the set of these equivalence classes for fixed $m$, and we use this partial order to study the topological closure of an equivalence class of UPBs.
In the case of four qubits, for $m=8,9,10$, we construct explicitly the Hasse diagram of this partial order.
\end{abstract}

\maketitle

\tableofcontents

\section{Introduction}

Multiqubit systems of quantum physics have been physically realized in recent years. Investigating the properties of multiqubit states is both physically and mathematically meaningful. In particular, the multiqubit positive-partial-transpose (PPT) entangled states can be constructed using multiqubit unextendible product bases (UPBs) \cite{bdm99}. UPBs indicate the quantum nonlocality in the discrimination of product states by local operations and classical communication (LOCC). Further, the three-qubit PPT entangled state obtained from a UPB is a biseparable entangled state, i.e., it is separable w.r.t. any bipartition of the three systems \cite{bravyi}. It exhibits a fundamental difference between the bipartite and multipartite quantum systems. 

We consider the UPBs of fixed cardinality $m$ in quantum systems of $n$ qubits. These UPBs are divided into finitely many equivalence classes with respect to an equivalence relation introduced by N. Johnston \cite{johnston14}. There is a natural partial order ``$\le$'' on the set of these equivalence classes for fixed $m$, and we use this partial order to study the topological closure of an equivalence class of UPBs.
A UPB is called {\em proper} if it does not span the whole 
Hilbert space of the system. 

Let $\cU$ and $\cU'$ be two $m$-state UPBs of an $n$-qubit system. We raise the question of deciding whether $\cU'$ lies in the closure of the equivalence class $\cE$ which contains $\cU$. 
Roughly speaking this means whether $\cU'$ can be approximated (to an arbitrary small precision) by UPBs belonging to $\cE$.
We solve this problem by using the above mentioned partial order, see Eq. \eqref{eq:closureUOM} and Proposition \ref{pp:closure}.

In the case of four qubits, it is known that the cardinality $m$ of an UPB is one of the numbers $6,7,8,9,10,12,16$. The construction of the proper UPBs (those with $m<16$) has been completed in 2014 by N. Johnston. He also classified these UPBs up to equivalence \cite{johnston14,johnston14c}. There are in total 1446 equivalence classes of proper UPBs of four qubits. For $m\le 10$ we describe the corresponding partial order 
``$\le$'' explicitly by constructing all arrows of its Hasse diagram. Due to a large number of equivalence classes, only a partial result is obtained for $m=12$. In all cases except 
$m=16$ we have determined the number of connected components of the Hasse diagram as well as the maximal and minimal equivalence classes.

In our recent paper \cite{cd18} we have introduced a novel 
method to study the $n$-qubit UPBs. In the next section we  recall this method, give the definitions and cite some facts that we need. In particular, we define there the unextendible orthogonal matrices (UOMs). Every $m\times n$ UOM, say $X$, generates an infinite family of $m$-state $n$-qubit UPBs which we denote by $\cF_X^\#$.
In Sec. \ref{sec:Order} we define the partial order ``$\le$'', 
introduce its Hasse diagram, recall the definition of 
decomposable orthogonal matrices and state some of their 
properties.
In Sec. \ref{sec:closure} we investigate the closure of equivalence classes of UPBs.
In Sec. \ref{sec:hasse} we describe the four-qubit Hasse diagrams of $m$-state UOMs for $m=8,9,10,12$, and identify the 
maximal and minimal equivalence classes of UOMs. 
In the Appendix \ref{app1} we list the representatives of the equivalence classes of UOMs of four qubits that we use. 
The arrows of the Hasse diagrams are listed in Appendix \ref{app2}.

\section{Preliminaries}
\label{sec:Prelim}

Let $\cH=\cH_1\ox\cdots\ox\cH_n$ be the Hilbert space representing a quantum system $A_1,\ldots,A_n$ consisting of $n$ qubits. 
Each $\cH_j$ is a 2-dimensional Hilbert space. We fix an 
orthonormal basis $\ket{0}_j,\ket{1}_j$ of $\cH_j$. Usually, 
the subscript $j$ will be suppressed. 
We say that a vector $\ket{v}\in\cH$ is a unit vector 
if $\|v\|=1$. For any nonzero vector $\ket{v_j}\in\cH_j$ 
we denote by $[v_j]$ the 1-dimensional subspace of 
$\cH_j$ spanned by this vector. As a rule, we shall not 
distinguish two unit vectors which differ only in the 
phase. By using this convention, we can say that for 
any unit vector $\ket{v_j}\in\cH_j$ there exists a unique 
unit vector $\ket{v_j}^\perp\in\cH_j$ which is perpendicular 
to $\ket{v_j}$. 

A {\em product vector} is a nonzero vector 
$\ket{x}=\ket{x_1}\ox\cdots\ox\ket{x_n}$, which will be 
written also as $\ket{x}=\ket{x_1,\ldots,x_n}$. 
If $\|x\|=1$ we shall assume (as we may) that each $\|x_j\|=1$. 
Two product vectors $\ket{x}=\ket{x_1,\ldots,x_n}$ and 
$\ket{y}=\ket{y_1,\ldots,y_n}$ are orthogonal if and only if 
$\ket{y_j}=\ket{x_j}^\perp$ for at least one index $j$. 
An {\em orthogonal product set (OPS )} is a set of pairwise 
orthogonal unit product vectors in $\cH$. 
The cardinality of an OPS cannot exceed $2^n$, the 
dimension of $\cH$. We say that an OPS is an {\em orthogonal product basis (OPB)}, if its cardinality is $2^n$.
As an example, the $2^n$ product vectors 
$\ket{x_s}=\ket{s_1,\ldots,s_n}$, where 
$s:=(s_1,\ldots,s_n)$ runs through all binary 
$\{0,1\}$-sequences of length $n$, is an OPB. 
We refer to this OPB as the {\em standard OPB}. 
However, there are many other OPBs. 
A set of unit product vectors is called an {\em unextendible product basis (UPB)} if these vectors are orthogonal to each other and there is no product vector orthogonal to all of them \cite{bdm99,DiV03}. 
Note that we allow the possibility that UPB spans the 
whole Hilbert space $\cH$. 

Our method uses formal matrices whose entries 
are vector variables represented by letters of an infinite countable alphabet $\bX$. 
Each $x\in\bX$ has its companion $x'\in\bX$ such 
that $x\ne x'$ and $(x')'=x$. We say that $x'$ is the 
{\em perpendicular} of $x$. 
We say that a subset of $\bX$ is {\em independent} if it does 
not contain any pair of the form $\{x,x'\}$. We say that two vector varables $x$ and $y$ are {\em independent} if 
$x\notin\{y,y'\}$. 

By $\cM(m,n)$ we denote the set of $m\times n$ matrices $X=[x_{ij}]$, $x_{ij}\in\bX$, such that if $x$ occurs in some column of $X$ then neither $x$ nor $x'$ occurs in any other column of $X$.

Let 
$x=[ ~ x_1 ~ x_2 ~ \cdots ~ x_n ~ ]$ and
$y=[ ~ y_1 ~ y_2 ~ \cdots ~ y_n ~ ]$ be two row vectors 
with $x_j,y_j\in\bX$. 
We say that $x$ and $y$ are {\em orthogonal} to each other, 
$x\perp y$, if $y_j=x'_j$ for at least one index $j$.

We say that a matrix $X\in\cM(m,n)$ is {\em orthogonal} if any two of its rows are orthogonal to each other. 
For two matrices $X$ and $Y$ with the same number of columns we say that they are {\em orthogonal to each other} $(X\perp Y)$ if each row of $X$ is orthogonal to each row of $Y$.
  
We denote by  $\cO(m,n)$ the subset of $\cM(m,n)$ consisting of all orthogonal matrices. We also set 
$\cO(n):=\cO(2^n,n)$ for the special case $m=2^n$.

The objects that we are interested in are the unextendible orthogonal matrices (UOM). We say that a matrix $X\in\cO(m,n)$ is {\em unextendible} if there is no row which is orthogonal to $X$. The reader can easily verify that the following four small matrices are UOMs: 
\begin{equation}
\label{eq:xy}
\left[\begin{array} {c}
a  \\
a' \\ \end{array} \right], \quad
\left[\begin{array} {cc}
a & b  \\
a & b' \\
a'& c  \\
a'& c' \end{array} \right], \quad
\left[\begin{array} {ccc}
a & c & e  \\
a'& d'& f  \\
b & c'& f' \\
b'& d & e' \end{array} \right], \quad
\left[\begin{array} {cccc}
x  & y  & z  & w  \\
x' & b  & d  & e  \\
a  & y' & d' & f  \\
a' & c  & z' & e' \\
a  & b' & d  & w' \\
x' & c' & d' & f' \end{array} \right].
\end{equation}

We say that two matrices $X,Y\in\cM(m,n)$ are 
{\em equivalent} if $X$ can be transformed to $Y$ by 
permuting the rows, permuting the columns, and by renaming 
the vector variables. The renaming must respect the orthogonality, i.e., we require that if a vector variable $x$ is renamed to $y$ then $x'$ has to be renamed to $y'$. For $X\in\cM(m,n)$ we shall denote by $[X]$ its equivalence class. 
Note that if $X\in\cO(m,n)$ then $[X]\subseteq\cO(m,n)$.
Since $\bX$ is infinite, there are infinitely many matrices in $\cM(m,n)$. On the other hand, there are only finitely many equivalence classes in $\cM(m,n)$. 

For example, for the UOM  
$X=\left[\begin{array}{c}a\\a'\end{array}\right]\in\cM(2,1)$ the equivalence class $[X]$ consists of all matrices 
$\left[\begin{array}{c}x\\x'\end{array}\right]$, $x\in\bX$.

The equivalence of two multiqubit UPBs has been defined in \cite[p. 4]{johnston14}. Let us recall that definition. Two multiqubit UPBs are {\em equivalent} if they have the same orthogonality graphs up to permuting the qubits and relabeling of vertices. For the definition of orthogonality graphs see \cite[Definition 4]{DiV03} or \cite[p. 3]{johnston14}.

There is a natural one-to-one correspondence between the equivalence classes of UOMs and the equivalence classes 
of UPBs. To explain this correspondence, we need the concept 
of evaluations. 

Given $X\in\cO(m,n)$, we define an {\em evaluation} $\alpha$ 
of $X$ to be a mapping which, for each $j=1,2,\ldots,n$,  replaces each vector variable $x$ in column $j$ by a unit vector $\alpha(x)$ in the 2-dimensional Hilbert space $\cH_j$ of the 
$j$th qubit. It is mandatory that $\alpha(x')=\alpha(x)^\perp$ whenever $x'$ also occurs in $X$. As a result of applying an evaluation $\a$ on $X$, we obtain an $m\times n$ matrix whose entries are unit vectors in the corresponding Hilbert spaces $\cH_j$. We denote this matrix by $\alpha(X)$. After that we can form $m$ product vectors by simply taking the tensor product of the unit vectors in a row of $\alpha(X)$. In this way we obtain an orthogonal set, $\cU$, of product vectors (OPS) in $\cH$. 
We refer to this OPS as the OPS of $\alpha(X)$.

In general, even when $X$ is a UOM, $\cU$ is not necessarily a UPB. To ensure that $\cU$ is a UPB, we have to require that if two independent vector variables, say $x$ and $y$, occur in the same column of $X$, then $\a(y)\ne\a(x)$ and 
$\a(y)\ne\a(x)^\perp$. For an arbitrary $X\in\cO(m,n)$, we say that an evaluation $\a$ of $X$ satisfying this additional condition is {\em generic}. It is proved in \cite[Lemma 2]{cd18} that if $X$ is a UOM and $\a$ is a generic evaluation of $X$ then the OPS of $\a(X)$ is in fact a UPB.

We denote by $\cF_X$ the set of all OPS of $\a(X)$ where $\a$ ranges over all evaluations of $X$, and by $\cF_X^\#$ we denote the set of all OPS of $\a(X)$ where $\a$ ranges only over all generic evaluations of $X$. 
If $X$ is a UOM then it is easy to see that any two UPBs in $\cF_X^\#$ are equivalent to each other according to the definition of equivalence given above.

The equivalence class of UPBs which corresponds to the equivalence class $[X]$ of a UOM $X$ is the equivalence class which contains the set $\cF_X^\#$. Explicitly, this equivalence 
class is the union
\begin{equation} 
\label{eq:classUOM}
\bigcup_{\s\in S_n} \cF_{X^\s}^\#
\end{equation}
where $\s$ runs over all permutations in the symmetric group $S_n$, and $X^\s$ is the UOM obtained by permuting the columns of $X$ by $\s$.

For $X\in\cM(m,n)$ we denote by $\nu_j(X)$ the cardinality of a maximal independent subset of the set of all entries in column $j$ of $X$. 
(All maximal independent subsets have the same cardinality.) 
We also set $\nu(X)=\sum_j \nu_j(X)$. If $\nu(X)=l$ we say 
that $X$ {\em lies on level $l$}.

We define a binary relation ``$\prec$'' on $\cM(m,n)$ which is 
a slight modification but equivalent to the definition given in \cite{cd18}. Let $X\in\cM(m,n)$ and $1\le j\le n$. Further, let 
$x,y$ be independent vector variables such that $x$ or $x'$ and $y$ or $y'$ occur in column $j$ of $X$. 
Denote by $Y$ the matrix obtained from $X$ by replacing all occurrencies (if any) of $x$ and $x'$ in $X$ by $y$ and $y'$, respectively. Then we write $Y\prec X$ and we say that $Y$ is obtained from $X$ by the {\em identification} $y=x$.
Note that $X$ and $Y$ differ only in column $j$ and that $\nu_j(Y)=\nu_j(X)-1$. We warn the reader that the expression 
``identification $y=x$'' does not mean that the vector variables $x$ and $y$ are the same, it just means that we are getting rid of the variables $y$ and $y'$ in $X$ and replacing them with $x$ and $x'$, respectively.

If $Y\prec X$ and $X$ is a UOM then $Y$ is orthogonal but 
does not have to be a UOM. For instance this is the case for 
\begin{equation}
\label{eq:notUOM}
X=\left[\begin{array} {ccc}
a & c & e  \\
a'& d'& f  \\
b & c'& f' \\
b'& d & e' \end{array} \right], \quad
Y=\left[\begin{array} {ccc}
a & c & e  \\
a'& d'& e  \\
b & c'& e' \\
b'& d & e' \end{array} \right].
\end{equation}
Indeed $Y\prec X$ by the identification $f=e$, and $X$ is a UOM while $Y$ is not as $[~a~c'~e~] \perp Y$.

The relation $\prec$ extends naturally to equivalence 
classes of UOMs. If $\cX,\cY$ are two equivalence classes
of UOMs, we write $\cY\prec\cX$ if $Y\prec X$ for some 
$X\in\cX$ and some $Y\in\cY$.

\section{Partial order}
\label{sec:Order}

If $X=[x_{i,j}]\in\cM(m,n)$ and $x\in\bX$ we define the 
{\em multiplicity}, $\mu(x,X)$, of $x$ in $X$ to be the number 
of pairs $(i,j)$ such that $x_{i,j}=x$. Thus if $x$ does not 
occur in $X$ then $\mu(x,X)=0$. When $X$ is known from the context we shall simplify this notation by writing just 
$\mu(x)$. Finally, we set 
$\mu(X)=\max_{i,j} \mu(x_{i,j},X)$.

Let $X\in\cM(m,n)$. If $\mu(x,X)=\mu(x',X)$ for all vector variables $x$ in column $j$ of $X$, then we say that the column 
$j$ of $X$ is {\em balanced}, and otherwise we say that it is {\em imbalanced}. We say that $X$ is {\em balanced} if each of its columns is balanced, and otherwise we say that $X$ is 
{\em imbalanced}. It is obvious that $X$ is imbalanced if $m$ is 
odd. We have shown in \cite[]{cd17JPA} that all UOM in $\cO(n)$ are necessarily balanced.

Next we recall from \cite{cd18} the definition of the partial order ``$\le$'' in $\cM(m,n)$. For two matrices $X,Y\in\cM(m,n)$, we say that $Y\le X$ if there exists a finite chain 
\begin{equation} \label{eq:chain}
Y=Z_0\prec Z_1\prec \cdots \prec Z_k =X,\quad k\ge 0. 
\end{equation}
We write $Y<X$ if $Y\le X$ and $Y\ne X$. Note that if 
$X\in\cO(m,n)$ then all the $Z_i$ in the chain \eqref{eq:chain} 
belong to $\cO(m,n)$. 
Further, if $X$ and $Y$ in \eqref{eq:chain} are UOM then so are all the $Z_i$. This follows immediately from 
\cite[Lemma 17]{cd18}.


Let us also recall the definition of maximal and minimal UOMs.

\begin{definition} \label{df:max}
We say that a UOM $X\in\cO(m,n)$ is {\em maximal} if there is no UOM $Y\in\cO(m,n)$ such that $X\prec Y$. Similarly, we say that a UOM $X\in\cO(m,n)$ is {\em minimal} if there is no UOM 
$Y\in\cO(m,n)$ such that $Y\prec X$. Further we say that a UOM is {\em isolated} if it is both maximal and minimal. 
\end{definition}

The definition of the partial order ``$\le$'' on $\cM(m,n)$  extends naturally to equivalence classes of matrices in $\cM(m,n)$. If $\cX$ and $\cY$ are two equivalence classes of matrices in $\cM(m,n)$ and $Y\le X$ for some $X\in\cX$ and some 
$Y\in\cY$, then we shall write $\cY\le\cX$. If $\cX\le\cY$ and $\cX\ne\cY$ then we write $\cX<\cY$. Further, we write 
$\cY\prec\cX$ if $Y\prec X$ for some $X\in\cX$ and some 
$Y\in\cY$.

We can also extend the definition of maximal and minimal UOM to the equivalence classes of UOMs. E.g. we say that an equivalence class $\cX\subseteq\cO(m,n)$ of UOMs is {\em maximal} if there is no equivalence class of UOMs $\cY\subseteq\cO(m,n)$ such that $\cX<\cY$. 

Next, we say that a UOM $X$ and its equivalence class $[X]$  are {\em reducible} if $\nu_j(X)=1$ for at least one $j$. Otherwise we say that $X$, and $[X]$, are {\em irreducible}. 

For convenience, we denote by $\UOM[m,n]$ the set of equivalence classes of UOMs in $\cO(m,n)$. This is a finite partially ordered set with partial order ``$\le$''. 

\begin{definition} \label{df:Hasse}
If $\cX,\cY\in\UOM[m,n]$ and $\cY\prec\cX$ we shall write 
$\cX\to\cY$ and refer to it as an {\em arrow}. The set $\UOM[m,n]$ equipped with all arrows that exist between its members is the Hasse diagram of the partially ordered set $(\UOM[m,n],\le)$.
\end{definition}

The Hasse diagram of $\UOM[m,n]$ can be viewed as a graph by ignoring the direction of arrows. We shall refer to the connected components of this graph also as the connected components of the Hasse diagram.

The following construction has been introduced in \cite{cd18}.
Let $A\in\cM(r,s)$ and $B\in\cM(m,n-s)$, $n>s$, and assume 
that any vector variables $x$ and $y$ of $A$ and $B$, 
respectively, are independent. Further, let $B$ be 
partitioned into blocks $B_k\in\cM(m_k,n-s)$, $k=1,\ldots,r$, 
$$
B=\left[ \begin{array}{c} B_1 \\ \vdots \\ B_r \end{array}
\right].
$$
Then we denote by $A\models(B_1,B_2,\ldots,B_r)$ 
the matrix $[~\tilde{A}~B~]\in\cM(m,n)$, where $\tilde{A}$ 
is the $m\times s$ matrix obtained from $A$ by replacing, 
for each $k$, the row $k$ of $A$ by $m_k$ copies of that row.
Note that if a vector variable $x$ occurs in one of the 
blocks $B_k$ then $x$ or $x'$ may occur in another block 
but necessarily in the same column.

For instance we have

\begin{equation} \label{eq:dekomp} 
\left[ \begin{array}{c}
x  \\
x' \\
\end{array} \right] \models \left(
\left[ \begin{array}{c}
a  \\
a' \\
\end{array} \right],
\left[ \begin{array}{c}
b  \\
b' \\
\end{array} \right] \right)= 
\left[ \begin{array}{cc}
x  & a  \\
x  & a' \\
x' & b  \\
x' & b' \\
\end{array} \right].
\end{equation}

It is easy to see that if $A$ and the $B_k$ are orthogonal matrices, then the matrix $A\models(B_1,B_2,\ldots,B_r)$ is also orthogonal. It is shown in \cite[Proposition 7]{cd18} that the matrix $A\models(B_1,B_2,\ldots,B_r)$ is a UOM if and only if 
$A$ and all the $B_k$ are UOMs.

\begin{definition} \label{df:decomposable}
We say that a matrix $X\in\cM(m,n)$ is {\em decomposable} 
if it is equivalent to a matrix $A\models(B_1,B_2,\ldots,B_r)$, 
$r\ge 1$. 
\end{definition}

Let $X$ be a decomposable UOM, say $X=A\models(B_1,\ldots,B_r)$
where $A\in\cO(r,s)$ and $B_k\in\cO(m_k,n-s)$, 
$k=1,2,\ldots,r$, are UOMs and $\sum m_k=m$. Then we can easily decide whether $X$ is maximal or minimal. The first lemma below 
is proved in \cite[Lemma 18]{cd18}.

\bl \label{le:maxUOM}
$X$ is maximal if and only if $A$ and all $B_k$ are maximal and no two of the blocks $B_k$ have a vector variable in common.
If $X$ is not maximal then there exists a UOM $Y$ 
such that $X\prec Y$ and $Y$ is obtained from $X$ by modifying 
a single column in either $A$ or just one of the blocks $B_k$.
\el

\bl \label{le:minUOM}
$X$ is minimal if and only if $A$ and all $B_k$ are minimal and for each $j\in\{1,2,\ldots,n-s\}$ all vector variables 
which occur in column $j$ of all the $B_k$s already occur in just one of the $B_k$s.
If $X$ is not minimal then there exists a UOM $Y$ 
such that $Y\prec X$ and $Y$ is obtained from $X$ by making 
a single identification in either $A$ or just one of the blocks $B_k$.
\el

We omit the proof of this lemma as it is similar to the proof 
of \cite[Lemma 18]{cd18}.

\section{The closure of equivalence classes of UPBs} 
\label{sec:closure}

For convenience, let us denote by ${\bf OPS}_m$ the set of 
OPSs of cardinality $m$ in $\cH$, and by ${\bf UPB}_m$ the set of UPBs of cardinality $m$ in $\cH$. As mentioned in the introduction, for $X\in\cO(m,n)$, we define
\begin{eqnarray*}
\cF_X &=& \{{\rm OPS~of}~\a(X):~\a~{\rm evaluation~of}~X\}, \\
\cF_X^\# &=& \{{\rm OPS~of}~\a(X):~\a~{\rm generic~evaluation~of}~X\}. \\
\end{eqnarray*}

For an UOM $X\in\cO(m,n)$, we have that 
$\cF_X\subseteq{\bf OPS}_m$ and $\cF_X^\#\subseteq{\bf UPB}_m$.
By \cite[Lemma 3]{cd18}, each $\cU\in{\bf UPB}_m$ belongs to some $\cF_X^\#$. Moreover, from the proof of that lemma it follows easily that $X$ can be recovered, uniquely up to row permutations and renaming of vector variables, from any 
$\cU\in\cF_X^\#$. Thus we have

\bpp \label{pp:partition}
When $X$ runs through the set of representatives of the equivalence classes of UOMs in $\cO(m,n)$, then the sets 
\eqref{eq:classUOM} form a partition of ${\bf UPB}_m$. 
\epp

Consequently, for UOMs $X$ and $Y$ we have $[X]=[Y]$ if and only if $\cF_{X^\s}^\#=\cF_Y^\#$ for some $\s\in S_n$. One can also show (see Corollary \ref{cr:FX=FY}) that $[X]=[Y]$ if and only if $\cF_{X^\s}=\cF_Y$ for some $\s\in S_n$. 

Since we are identifying two unit product vectors if they differ only in phase, such product vectors are in fact points in the 
projective space $P\cH$ associated with $\cH$. More precisely, 
these points lie on the Segre subvariety 
$\S_n:=P\cH_1\times\cdots\times P\cH_n$ of $P\cH$. Consequently, 
after ordering the product vectors of a $\cU\in{\bf OPS}_m$, we obtain a point in $\S_n^m$, the product of $m$ copies of $\S_n$. As an example, if $\{\ket{\psi_i}:i=1,2,3,4\}$ is an OPS 
in $\cH$ then the ordered quadruple 
$\left( \ket{\psi_1},\ket{\psi_2},\ket{\psi_3},\ket{\psi_4} \right)$ can be viewed as a point of $\S_n^4$. 
Since the product vectors of $\cU$ can be ordered in 
$m!$ ways and they are orthogonal to each other, we obtain in fact $m!$ different points of $\S_n^m$. 

The set ${\bf OPS}_m$ is closed in $\S_n^m$ because it is defined by the equations saying that its product vectors are 
orthogonal to each other. However, its subset ${\bf UPB}_m$ is not closed in general. For instance, in the case $m=4$, $n=3$ consider the set $\cU_t$ of the four pure product states 
\begin{eqnarray*}
\ket{\psi_1(t)}&=& \ket{0} \ox \ket{0} \ox \ket{0}, \\
\ket{\psi_2(t)}&=& \ket{1} \ox \ket{e_{-}} \ox 
        \left( \cos t\,\ket{0}+\sin t\,\ket{1} \right), \\
\ket{\psi_3(t)}&=& \ket{e_{+}} \ox \ket{1} \ox 
        \left( -\sin t\,\ket{0}+\cos t\,\ket{1} \right), \\
\ket{\psi_4(t)}&=& \ket{e_{-}} \ox \ket{e_{+}} \ox \ket{1},
\end{eqnarray*}
where $\ket{e_{\pm}}=(\ket{0} \pm \ket{1}) / \sqrt{2}$, and 
$t$ is a real parameter. For $t\in(0,\pi/2)$ the set $\cU_t$ is 
a UPB while for $t=0$ it is an OPS which is not a UPB. As 
$\cU_0=\lim_{t\to 0} \cU_t$, we see that $\cU_0$ belongs to the 
closure of ${\bf UPB}_4$.

Our main goal in this section is to describe the closure of any equivalence class of $m$-state UPBs viewed as a subset of the variety $\S_n^m$. We first make the following observation. 

\bl \label{le:open}
For any $X\in\cO(m,n)$, the set $\cF_X^\#$ is open in $\cF_X$ 
(with respect to the relative topology of $\cF_X$).
\el
\bpf
Let $(x,y)$ be a pair of independent vector variables which occur in the same column of $X$. Denote by $\cF_X^{(x,y)}$ the subset of $\cF_X$ consisting of the OPS of $\a(X)$ where $\a$ 
runs through all evaluations of $X$ subject to the condition 
that $\a(y)=\a(x)$ or $\a(y)=\a(x)^\perp$. Clearly, this is a closed subset of $\cF_X$. The assertion now follows from the observation that 
$$
\cF_X^\#=\cF_X \setminus \bigcup_{(x,y)} \cF_X^{(x,y)},
$$
where $(x,y)$ runs through the finite set of all pairs of the kind mentioned above.
\epf

\bt \label{thm:closure}
If $X\in\cO(m,n)$ then $\overline{\cF_X^\#}=\cF_X$. 
\et
\bpf
First we note that the set $\cF_X$ is a closed subset of 
$\S_n^m$. This follows from the observation that $\cF_X$, 
considered as a subset of $\S_n^m$, is the image of a direct product of a finite number of copies of the projective spaces 
$P\cH_j$, $1\le j\le n$, under a continuous map to $\S_n^m$. 
Indeed let us select a maximal set $V$ of independent vector 
variables that occur in the matrix $X$. If $x\in V$ occurs in column $j$ of $X$, we denote by $S_x$ a copy of $P\cH_j$. 
Finally we set $S:=\times_{x\in V} S_x$. To any point 
$v:=(\ket{v_x})_{x\in V}\in S$ we attach an evaluation $\a_v$ of $X$ as follows: if $x\in V$ then we set $\a_v(x)=\ket{v_x}$ and 
if $x'$ occurs in $X$ we also set $\a_v(x')=\ket{v_x}^\perp$. We can now define a continuous map $f:S\to\S_n^m$ as follows: 
$f(v)$ is the OPS of the matrix $\a_v(X)$.
The image of $f$ is precisely the subset $\cF_X$ of $\S_n^m$. 
Since $S$ is compact, its image $f(S)=\cF_X$ is closed in $\S_n^m$. 
As $\cF_X^\#\subseteq\cF_X$, it follows that 
$\overline{\cF_X^\#}\subseteq\cF_X$. 

We shall now prove the opposite inclusion 
$\cF_X\subseteq\overline{\cF_X^\#}$. 
Let $\cU\in\cF_X$ be arbitrary. Then $\cU$ is the OPS of the matrix $\a(X)$ for some evaluation $\a$ of $X$. It is easy to see that there exists 
a finite chain \eqref{eq:chain} such that $\cU$ is the OPS of $\b(Y)$ where the evaluation $\b$ of $Y$ is the restriction of $\a$. Therefore it suffices to prove the assertion in the case 
where $Y\prec X$. 
Thus $Y$ can be obtained from $X$ by an identification $x=y$ where $x$ and $y$ are independent vector variables such that 
$x$ and at least one of $y$ and $y'$ occur in the same column of $X$, say in column $j$. We choose a continuous path 
$\s:[0,1]\to P\cH_j$ such that $\s(0)=\a(x)$ and 
$\s(1)=\a(y)$, and moreover $\s(t)\ne\a(z),\a(z)^\perp$ for all $t\in(0,1)$ and all vector variables $z$ in column $j$ of $X$. 
Let us now define a continuous one-parameter family $\a_t$, 
$0\le t\le1$, of evaluations of $X$ by setting 
$\a_t(x)=\s(t)$, $\a_t(x')=\s(t)^\perp$, and $\a_t(z)=\a(z)$ 
for all $z\ne x,x'$. Then $\a_0=\a$, $\a_1=\beta$ and 
$\a_t$ is generic for $0\le t<1$. Thus $\a_t(X)\in\cF_X^\#$ for $0\le t<1$. Clearly we have 
$\b(Y)=\lim_{t\to 1} \a_t(X)$. Since this limit belongs to the closure of $\cF_X^\#$, our assertion is proved.
\epf

\bcr \label{cr:FX=FY}
For $X,Y\in\cO(m,n)$, the equalities $\cF_X^\#=\cF_Y^\#$ and $\cF_X=\cF_Y$ are equivalent to each other. 
\ecr
\bpf
Theorem \ref{thm:closure} shows that the first equality implies the second one. We shall prove the converse. 

Assume that $\cF_X=\cF_Y$. By Lemma \ref{le:open}, $\cF_X^\#$ is open in $\cF_X$ and by the theorem it is also dense in $\cF_X$.
Hence, both $\cF_X^\#$ and $\cF_Y^\#$ are dense open subsets of $\cF_X$. Therefore $\cF_X^\#$ and $\cF_Y^\#$ cannot be 
disjoint, i.e., there exists $\cU\in\cF_X^\#\cap\cF_Y^\#$. As mentioned in the beginning of this section, $Y$ can be obtained from $X$ by permuting the rows and renaming the vector variables. Consequently $\cF_X^\#=\cF_Y^\#$.
\epf

Now let $X\in\cO(m,n)$ be any UOM. The corresponding equivalence class of $m$-state UPBs in $\cH$ is given by the expression 
\eqref{eq:classUOM}. As explained above, we can consider this equivalence class as a subset of $\S_n^m$. 
It follows immediately from Theorem \ref{thm:closure} that the closure of this equivalence class is the union
\begin{equation} 
\label{eq:closureUOM}
\bigcup_{\s\in S_n} \cF_{X^\s}.
\end{equation}

\bpp \label{pp:closure}
For $X,Y\in\cO(m,n)$ we have 

(i) $Y\le X \Longrightarrow \cF_Y \subseteq \cF_X;$

(ii) $Y<X\Longrightarrow\cF_Y\subseteq\cF_X\setminus\cF_X^\#.$
\epp
\bpf
(i) Since there is a chain \eqref{eq:chain} from $Y$ to $X$, 
it suffices to prove the assertion in the case where $Y\prec X$. Then $Y$ can be obtained from $X$ by performing a single identification, say $y=x$. For $\cU\in\cF_Y$ there exists an 
evaluation $\b$ of $Y$ such that $\cU$ is the OPS of the matrix 
$\b(Y)$. We can extend $\b$ to an evaluation $\a$ of $X$ by 
setting $\a(y)=\b(x)$ and $\a(y')=\b(x)^\perp$. Then $\cU$ is also the OPS of the matrix $\a(X)$, i.e., we have $\cU\in\cF_X$.

(ii) We shall use again the chain \eqref{eq:chain}. As $Y<X$ the length $k$ of this chain is at least 1. Thus we have 
$Y\le Z_{k-1}$. By using (i) we deduce that 
$\cF_Y\subseteq\cF_{Z_{k-1}}$. Hence, it suffices to prove (ii) 
in the case where $Y\prec X$. By (i) we know that 
$\cF_Y \subseteq \cF_X.$ It remains to show that 
$\cF_Y \cap \cF_X^\#=\emptyset.$

Let $\cU\in\cF_Y$ be arbitrary. Then $\cU$ is the OPS of $\b(Y)$ for some evaluation $\b$ of $Y$. Since $Y\prec X$, there is an  identification $y=x$ which transforms $X$ into $Y$. Let $\a$ be any generic evaluation of $X$. Since $x$ and $y$ are independent vector variables, we must have $\a(y)\ne\a(x)$. Consequently $\b(Y)\ne\a(X)$, and so $\cU\notin\cF_X^\#$. 
\epf

\section{The four-qubit Hasse diagrams}
\label{sec:hasse}

We shall construct all the arrows of the Hasse diagram of $\UOM[m,4]$ for $m=8,9,10$. In the case $m=12$ we construct only 
the arrows connecting the irreducible equivalence classes. 
We shall also determine the maximal and minimal equivalence classes of UOMs in all four cases $m=8,9,10,12$.  

The cases $m=6,7$ are omitted because each of them contains only one equivalence class. 
The case $m=16$ is also omitted though for a different reason. In that case the equivalence classes have not been enumerated so far, only the maximal equivalence classes are known \cite{cd17JPA}.

In the table below we show the distribution of the equivalence classes of UOMs in $\cO(m,4)$ over various levels 
$\nu=7,8,\ldots,14$.

\begin{table}[h] 
$$
\begin{array}{ccccccc}
\nu\backslash m & 6 & 7 & 8 & 9 & 10 & 12 \\
\hline
14 &   &   &    &   &    &   2 \\
13 &   &   &  1 &   &    &  11 \\
12 &   &   &  1 &   &    &  46 \\
11 &   &   &  6 &   &  2 & 154 \\
10 & 1 &   & 29 & 1 & 17 & 332 \\
 9 &   & 1 & 46 & 4 & 37 & 392 \\
 8 &   &   & 43 & 6 & 24 & 227 \\
 7 &   &   & 18 &   &    &  45 \\
\hline
{\rm Total} &1&1 &144 &11 & 80 &1209 \\
\end{array}
$$
\caption{The distribution of the equivalence classes of UOMs in $\cO(m,4)$ over various levels.}
\end{table}

For a fixed value of $m$, the classes on the top level are maximal and those on the bottom level are minimal. For instance, in $\cO(10,4)$ both classes on level 11 are maximal and all 24 
classes on level 8 are minimal.

Most of the proofs rely on the results of 
Johnston \cite{johnston14}. In particular we have used his 
classification of four qubit UPBs in order to construct the
above table.

For a given UOM $X$, we refer to the quadruple 
$[\n_1(X),\ldots,\n_4(X)]$ as $\n$-{\em numbers} of $X$. 
We shall also use the $\mu$-{\em numbers} $[\m_1(X),\ldots,\m_4(X)]$ of $X$, where $\m_j(X)$ is the largest multiplicity of the vector variables in column $j$ of $X$.

In Appendix \ref{app1} we have listed the representatives of the equivalence classes in $\UOM[m,4]$ for $m=8,9,10,12$. For 
$m=12$ we list only the representatives of the irreducible classes. These lists are extracted from \cite{johnston14c}. 
(They are presented in a different format, suitable for this paper.) Our listing is by levels, starting from the top level and ending with the bottom level. We record one matrix per line by using the following conventions:

(i) the rows are listed one after the other and separated by white space;

(ii) we only use the vector variables $a_j,b_j,c_j,d_j$, 
$j=1,2,3,4$, and their perpendiculars which we denote here by 
capital letters $A_j,B_j,C_j,D_j$, respectively;

(iii) we omit the subscripts as they can be easily 
recovered: each vector variable in column $j$ should have the 
subscript $j;$

(iv) we end each line by listing the $\n$-numbers of the matrix.

We denote by $X_{l,k}$ the $k$th matrix in the list of representatives on level $l$. Note that the number, $m$, of rows of the UOM is supressed, it does not appear in the symbol $X_{l,k}$.

As an example, in the case $m=8$, the last matrix on level 11 is
$$ 
X_{11,6}=
\left[ \begin{array}{cccc}
a_1 & c_2 & c_3 & a_4 \\
a_1 & c'_2 & a_3 & c_4 \\
a_1 & a_2 & c'_3 & c'_4 \\
b_1 & a'_2 & a'_3 & a'_4 \\
a'_1 & b_2 & b_3 & a_4 \\
a'_1 & b'_2 & a_3 & b_4 \\
a'_1 & a_2 & b'_3 & b'_4 \\
b'_1 & a'_2 & a'_3 & a'_4 \\
\end{array} \right]
$$
and its $\n$-numbers are $[2,3,3,3]$. Their sum is $11$ and so 
this matrix lies on level 11. This is recorded as the first subscript of $X_{11,6}$. The second subscript $6$ means that this matrix occupies the sixth place in our list of representatives on level 11. 

In view of the large number of equivalence classes of UOMs, we 
had to examine many cases to find the arrows of the Hasse diagram. For that purpose we used computer programs 
that we wrote in Maple. We shall only sketch here the main 
steps of the search. 

For a given UOM say $X=X_{l,r}$ on level $l$ there are only finitely many matrices $Y$ such that $Y\prec X$. A computer program can easily generate all such matrices $Y$ and it 
can also test whether $Y$ is a UOM. If the test is negative, then $Y$ is discarded. Otherwise $Y$ is a UOM and we use another computer program which tests whether $Y$ is equivalent to some UOM on level $l-1$. In fact we know that this must be the case, 
there is a unique $s$ such that $Y$ is equivalent to $X_{l-1,s}$. Then we have found the arrow $X_{l,r}\to X_{l-1,s}$. The latter program is described in 
\cite[Section 4]{cd18}.  Unfortunately when $m=12$ or $m=10$ this program fails in many cases because it uses too much time. Such cases have to be dealth with separately. 
After processing in this way all UOMs on level $l$, we can decide which UOMs on level $l-1$ are maximal and which UOMs on level $l$ are minimal. 

The Appendix \ref{app2} consists of four subsections, one for each of the cases $m=8,9,10,12$. Each subsection contains several tables where we record the arrows of the Hasse diagram of $\UOM[m,4]$. Each of these tables contains the list of all 
arrows from level $l$ to level $l-1$ for fixed $l$.
For instance in the case $m=8$ the first table lists the arrows from level 11 to 10. The second line of that table is:
$$
2\to 4,5 \quad\quad  d_4=b_4,d_4=B_4 
$$
The meaning of $2\to 4,5$ is that we have two arrows: $2\to 4$ and $2\to 5$. More precisely, these arrows are 
$X_{11,2}\to X_{10,4}$ and $X_{11,2}\to X_{10,5}$ and they belong to the Hasse diagram of $\UOM[8,4]$ since $m=8$. 
The rest of that line contains two identifications: 
$d_4=b_4$ and $d_4=B_4$. These identifications allow us to verify the existence of the two arrows mentioned above. The 
first identification $d_4=b_4$ justifies the first arrow 
$2\to 4$, and $d_4=B_4$ justifies $2\to 5$. After performing the identification $d_4=b_4$ on the matrix $X_{11,2}$ we obtain 
say the matrix $Y$. The verification of the arrow $2\to 4$ is completed by showing that $Y$ is equivalent to $X_{10,4}$.

We now state the main results of our computations for each 
of the cases $m=8,9,10,12$.

\subsection{$\UOM[8,4]$} \label{subsection:(8,4)}

$\UOM[8,4]$ has cardinality 144 and occupies the levels $7-13$. 
We start with some facts that can be proved directly without using the computer. Let us first show that there is an arrow 
from the class represented by $X_{13,1}$ to the one represented by $X_{12,1}$. Note that these two matrices have the same 
first column, and that the columns 2 and 3 of $X_{13,1}$ are 
the same as the columns 3 and 4 of $X_{12,1}$, respectively. 
Next we identify the variable $c_4$ to $a_4$, i.e., we set 
$c_4=a_4$ and $C_4=A_4$, in the last column of $X_{13,1}$. 
(For the sake of clarity we use here subscripts which were 
omitted in the tables of Appendix \ref{app1}.) It is easy now to verify that this new matrix, let us call it $Y$, is 
equivalent to $X_{12,1}$. We just have to rename $d_4$ to $c_4$ 
(and $D_4$ to $C_4$) in the last column of $Y$ and then by permuting the columns we obtain the matrix $X_{12,1}$. 
We conclude that $X_{13,1}$ is not minimal and $X_{12,1}$ 
is not maximal.

Similarly one can verify that we have arrows 
$X_{12,1}\to X_{11,k}$ for $k=1,2,3,4,5$. 
On the other hand we claim that there is no arrow from $X_{12,1}$ to $X_{11,6}$. To prove this claim, it suffices to inspect their $\nu$ numbers, $[1,3,4,4]$ and $[2,3,3,3]$, respectively. If $Y\prec X_{12,1}$ is a UOM, then 3 of the 
$\nu$-numbers of $Y$ must be the same as the corresponding 
$\nu$-numbers of $X_{12,1}$. It follows that $Y$ and $X_{11,6}$ have different $\nu$-numbers, and so they are not equivalent. This proves our claim. We conclude that $X_{12,1}$ is not minimal, the UOMs $X_{11,k}$ are not maximal for $k=1,2,3,4,5$, and that $X_{11,6}$ is maximal.

The arrows from level $l=11$ to level 10 are listed in the first 
table of Appendix \ref{app2}.
The first line of that table indicates that there are four 
arrows emanating from the first UOM on level 11. Namely there
exist arrows from $X_{11,1}$ to $X_{10,k}$, $k=1,2,3,6$. 
The existence of the first of these arrows can be verified as 
follows. First we identify the independent variables $d_4$ and 
$a_4$ in column 4 of $X_{11,1}$ to obtain a matrix, say $Y$.
This means that we have to replace $d_4$ and $D_4$ with $a_4$ 
and $A_4$, respectively. After this is done we have to verify  
that $Y$ is equivalent to $X_{10,1}$. In this case this is very easy, we just switch the columns 3 and 4 of $Y$ to obtain the matrix $X_{10,1}$. Hence they are equivalent. 
(In this case one can identify two independent variables in 
the third column of $X_{11,1}$, to obtain another proof that the arrow $X_{11,1}\to X_{10,1}$ exists.)
For the remaining three arrows emanating from $X_{11,1}$ 
we use the identifications $d_4=B_4$, $d_4=A_4$, $d_4=b_4$,
respectively.

The following proposition follows from the above comments and
the list of arrows given in Appendix \ref{app2} for $m=8$.

\bpp \label{pp:(8,4)}
$\UOM[8,4]$ has cardinality $144$ and occupies the levels $7$ to $13$. There are $16$ maximal and $23$ minimal classes. 

The representatives of the maximal classes are 
$X_{13,1}$, $X_{11,6}$, $X_{10,[17-19]}$, 
$X_{10,[22-29]}$, $X_{9,46}$, $X_{8,24}$ and $X_{8,30}$.

The minimal classes are the 18 classes on level 7 and the classes with representatives $X_{8,25}$, $X_{8,34}$ and $X_{9,[26-28]}$.

\begin{figure}
\setlength{\unitlength}{1.2 mm}
\begin{center}
\begin{picture}(50,30)(-32,0)

\put(-50,24){$17$}
\put(-48,22){\circle*{1.2}}
\put(-48,22){\vector(0,-1){9}}
\put(-48,12){\circle{1.2}}
\put(-50,8){$26$}

\put(-40,24){$18$}
\put(-38,22){\circle*{1.2}}
\put(-38,22){\vector(0,-1){9}}
\put(-38,12){\circle{1.2}}
\put(-40,8){$27$}

\put(-28,22){\vector(-1,-1){9}}

\put(-30,24){$19$}
\put(-28,22){\circle*{1.2}}
\put(-28,22){\vector(0,-1){9}}
\put(-28,12){\circle{1.2}}
\put(-30,8){$28$}

\put(-23,10.5){$29$}
\put(-18,12){\circle{1.2}}

\put(-8,22){\vector(-1,-1){9}}
\put(2,22){\vector(-1,-1){9}}

\put(-18,12){\vector(1,-1){9}}
\put(2,22){\vector(1,-1){9}}
\put(-8,12){\vector(1,-1){9}}
\put(-8,22){\vector(1,-1){9}}

\put(-10,24){$23$}
\put(-8,22){\circle*{1.2}}
\put(-8,22){\vector(0,-1){9}}
\put(-8,12){\circle{1.2}}
\put(-13,10.5){$34$}
\put(-8,12){\vector(0,-1){9}}
\put(-8,2){\circle{1.2}}
\put(-10,-2){$25$}

\put(0,24){$22$}
\put(2,22){\circle*{1.2}}
\put(2,22){\vector(0,-1){9}}
\put(2,12){\circle{1.2}}
\put(3.5,10.5){$35$}
\put(2,12){\vector(0,-1){9}}
\put(2,2){\circle{1.2}}
\put(0,-2){$34$}

\put(12,12){\circle{1.2}}
\put(13.5,10.5){$36$}

\put(12,12){\vector(-2,-1){19}}

\put(30,21){$\n=10$}
\put(30,11){$\n=9$}
\put(30,1){$\n=8$}

\end{picture}
\end{center}
\caption{$\UOM[8,4]$, Hasse diagrams of the three small connected components. \\ The maximal classes are marked by bullets.} 
\label{fig:3connected}
\end{figure}
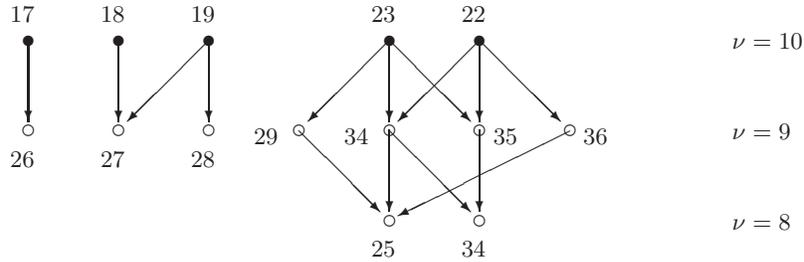

The Hasse diagram has four connected components. The three small 
components (shown on Fig. \ref{fig:3connected}) have as vertices the classes of 

\begin{eqnarray*}
&& X_{10,17},X_{9,26}; \\
&& X_{10,18},X_{10,19},X_{9,27},X_{9,28}; \\
&& X_{10,[22-23]},X_{9,29},X_{9,[34-36]},X_{8,25},X_{8,34},
\end{eqnarray*}
respectively. 
\epp

\subsection{$\UOM[9,4]$}

$\UOM[9,4]$ has cardinality 11 and occupies only the levels 8-10. 

\bpp \label{pp:(9,4)}
The Hasse diagram of $\UOM[9,4]$ has exactly $10$ arrows, see Appendix \ref{app2} and Fig. \ref{fig:o94}. It has two connected components, one of them consists of the isolated class  $[X_{8,6}]$. 
The representatives of the maximal classes are: 
$X_{10,1}$, $X_{9,1}$, $X_{9,2}$ and $X_{8,6}$. 
The minimal classes are all $6$ classes on level $8$.
\epp

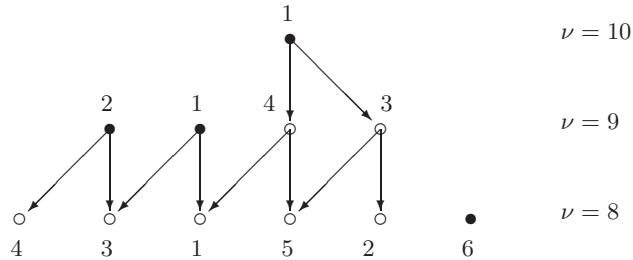
\begin{figure}
\setlength{\unitlength}{1.2 mm}
\begin{center}
\begin{picture}(50,32)(-12,-2)

\put(-10,10){\vector(-1,-1){9}}
\put(-10,10){\vector(0,-1){9}}
\put(0,10){\vector(-1,-1){9}}
\put(0,10){\vector(0,-1){9}}
\put(10,10){\vector(-1,-1){9}}
\put(10,10){\vector(0,-1){9}}
\put(20,10){\vector(-1,-1){9}}
\put(20,10){\vector(0,-1){9}}
\put(10,20){\vector(1,-1){9}}
\put(10,20){\vector(0,-1){9}}

\put(-20,0){\circle{1.2}}
\put(-10,0){\circle{1.2}}
\put(0,0){\circle{1.2}}
\put(10,0){\circle{1.2}}
\put(20,0){\circle{1.2}}
\put(30,0){\circle*{1.2}}

\put(-10,10){\circle*{1.2}}
\put(0,10){\circle*{1.2}}
\put(10,10){\circle{1.2}}
\put(20,10){\circle{1.2}}

\put(10,20){\circle*{1.2}} 

\put(9,22){$1$}
\put(-11,12){$2$}
\put(-1,12){$1$}
\put(7,12){$4$}
\put(20,12){$3$}

\put(-21,-4){$4$}
\put(-11,-4){$3$}
\put(-1,-4){$1$}
\put(9,-4){$5$}
\put(18,-4){$2$}
\put(29,-4){$6$}

\put(40,20){$\n=10$}
\put(40,10){$\n=9$}
\put(40,0){$\n=8$}

\end{picture}
\end{center}
\caption{$\UOM[9,4]$, Hasse diagram of equivalence classes. This is the whole Hasse diagram.} 
\label{fig:o94}
\end{figure}

\subsection{$\UOM[10,4]$}

$\UOM[10,4]$ has cardinality 80 and occupies the levels 8-11.

\bpp \label{pp:10:4}
$\UOM[10,4]$ has exactly $14$ maximal equivalence classes. Their representatives are $X_{11,1}$ and $X_{11,2}$ on level 
$11$, $X_{10,8-17}$ on level $10$, and $X_{9,17}$ and $X_{9,24}$ on level $9$. The minimal equivalence classes are all 
$24$ classes on level $8$. The Hasse diagram has two connected components, the smaller component consisting of $20$ classes is sketched on Figure \ref{fig:SMALLconnected}.
\epp
\bpf
All the assertions follow from the tables in Appendix \ref{app2} for $m=10$ where all arrows of the Hasse diagram are listed.
\epf

\begin{figure}
\setlength{\unitlength}{1.2 mm}
\begin{center}
\begin{picture}(50,32)(-22,0)

\put(-50,14){$17$}
\put(-48,12){\circle*{1.2}}
\put(-48,12){\vector(0,-1){9}}
\put(-48,2){\circle{1.2}}
\put(-50,-2){$11$}

\put(-42,14){$20$}
\put(-38,12){\circle{1.2}}
\put(-38,12){\vector(0,-1){9}}
\put(-38,2){\circle{1.2}}
\put(-40,-2){$13$}

\put(-33,12){$21$}
\put(-28,12){\circle{1.2}}
\put(-28,12){\vector(0,-1){9}}
\put(-28,2){\circle{1.2}}
\put(-30,-2){$14$}

\put(-13,13){$27$}
\put(-8,12){\circle{1.2}}
\put(-8,12){\vector(0,-1){9}}
\put(-8,2){\circle{1.2}}
\put(-10,-2){$16$}

\put(9,15){$30$}
\put(12,12){\circle{1.2}}
\put(12,12){\vector(0,-1){9}}
\put(12,2){\circle{1.2}}
\put(10,-2){$22$}

\put(17,13){$29$}
\put(22,12){\circle{1.2}}
\put(22,12){\vector(0,-1){9}}
\put(22,2){\circle{1.2}}
\put(20,-2){$23$}

\put(-30.5,25){$10$}
\put(-28,22){\circle*{1.2}}
\put(-28,22){\vector(0,-1){9}}
\put(-28,2){\circle{1.2}}

\put(-10.5,25){$16$}
\put(-8,22){\circle*{1.2}}
\put(-8,22){\vector(0,-1){9}}
\put(-8,2){\circle{1.2}}

\put(0,25){$12$}
\put(2,22){\circle*{1.2}}
\put(2,22){\vector(0,-1){9}}
\put(2,12){\circle{1.2}}
\put(-2.5,13){$28$}
\put(2,2){\circle{1.2}}
\put(-0.5,-2){$18$}

\put(19.5,25){$13$}
\put(22,22){\circle*{1.2}}
\put(22,22){\vector(0,-1){9}}
\put(22,2){\circle{1.2}}

\put(-23,12){$23$}
\put(-18,12){\circle{1.2}}
\put(32,12){\circle{1.2}}
\put(34,11){$31$}
\put(-28,22){\vector(-1,-1){9}}
\put(-8,22){\vector(-1,-1){9}}
\put(2,22){\vector(-1,-1){9}}
\put(22,22){\vector(-1,-1){9}}

\put(-18,12){\vector(-1,-1){9}}

\put(2,12){\vector(-1,-1){9}}
\put(12,12){\vector(-1,-1){9}}

\put(-48,12){\vector(1,-1){9}}
\put(-38,12){\vector(1,-1){9}}
\put(22,22){\vector(1,-1){9}}

\put(-8,22){\vector(-2,-1){19}}
\put(22,12){\vector(-2,-1){19}}
\put(-28,12){\vector(2,-1){19}}
\put(-18,12){\vector(2,-1){19}}
\put(-8,12){\vector(2,-1){19}}

\put(-8,22){\vector(2,-1){19}}

\put(32,12){\vector(-3,-1){29}}

\put(46,21){$\n=10$}
\put(46,11){$\n=9$}
\put(46,1){$\n=8$}

\end{picture}
\end{center}
\caption{$\UOM[10,4]$, Hasse diagram of the small connected component. There are two connected components, the one shown is the smaller one. The bigger component has $60$ vertices and we omit the picture.} 
\label{fig:SMALLconnected}
\end{figure}
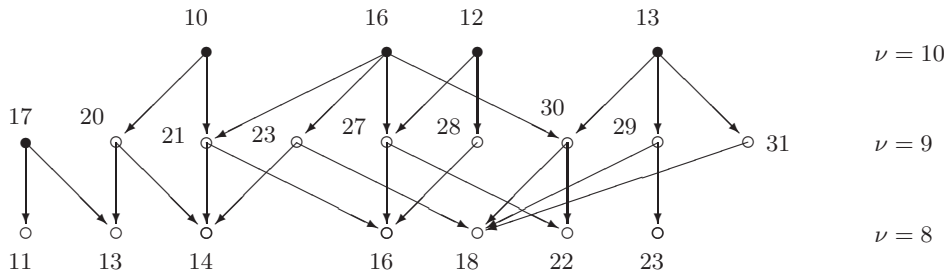

\subsection{$\UOM[12,4]$}

$\UOM[12,4]$ has cardinality $1209$ and occupies the levels $7-14$. Among the $1209$ classes only $161$ are irreducible. The latter occupy the levels $8-12$. There are 26,64,51,18,2 irreducible equivalence classes on these levels, respectively. 
For each of these levels $l$, the representatives of the irreducible equivalence classes are listed in Appendix \ref{app1} for $m=12$. 

\bpp \label{pp:12:4}
We consider here the equivalence classes of UOMs in $\cO(12,4)$.

(i) There are exactly $45$ minimal classes, namely all the classes lying on level $7$.

(ii) There are exactly $11$ maximal classes. Three of them are reducible, two on level $14$ and one on level $13$. The eight irreducible maximal classes have as representatives the UOMs: $X_{12,1}$ and $X_{12,2}$ on level $12$; $X_{11,1}$, $X_{11,2}$, $X_{11,17}$ and $X_{11,18}$ on level $11$; and $X_{10,50}$ and $X_{10,51}$ on level $10$. 
\epp
\bpf
Let $X\in\cO(12,4)$ be a reducible UOM. Up to equivalence, we 
may assume that 
$$
X=\left[\begin{array}{c}x\\x'\end{array}\right]\models(U,V),
$$
where $U$ and $V$ are UOMs of 3 qubits.
Since the UPBs of 3 qubits have cardinality 4 or 8, we may 
assume that $U\in\cO(8,3)$ and $V\in\cO(4,3)$. Note that $V$ 
is unique up to equivalence and that its $\nu$-numbers are 
$[2,2,2]$.

(i) Assume now that the above reducible UOM $X$ is minimal. 
By Lemma \ref{le:minUOM}, $U$ must be minimal.
By \cite[Corollary 15 and Figure 1]{cd17JPA}, $U$ is unique up to equivalence and its $\nu$-numbers are $[1,1,1]$. By the same lemma, each vector variable in $U$ must occur in $V$. Hence the $\nu$-numbers of $X$ are $[1,2,2,2]$ and so $\nu(X)=7$. Since there are no UOMs on level 6, all UOMs on level 7 are minimal. 

The tables in Appendix \ref{app2} show that there are no irreducible minimal UOMs on levels higher than 8, and by using a computer program we have verified that the same is true for 
irreducible UOMs on level 8. This completes the proof of (i).

(ii) Let $X$ be a reducible UOM given by the above formula and assume that it is maximal. Then, by Lemma \ref{le:maxUOM},  
$U$ and $V$ must be maximal and they have no vector variable in common. Hence, we can assume that $U$ is a representative of one of the three maximal equivalence classes of OPBs in $\cO(8,3)$ (see \cite[Fig. 1]{cd17JPA}). Two of them, classes 1 and 2, have
$\n(U)=7$ and the third one, class 6, has $\n(U)=6$. 
By permuting the last four rows of $X$, we may also assume 
that $V$ is the third matrix displayed in \eqref{eq:xy}. 
Since $\n(X)=1+\n(U)+\n(V)=\n(U)+7$, we have $\n(X)=14$ 
in the first two cases and $\n(X)=13$ in the third case.

Note that there are no arrows from a reducible UOM to an irreducible UOM. Hence, the assertion about the irreducible 
maximal equivalence classes follows from the tables in Appendix \ref{app2}. 
\epf

Lemma \ref{le:maxUOM} (ii) implies that all reducible classes of UOMs in $\cO(12,4)$ can be obtained from just the 3 maximal ones by applying the relation ``$\prec$''.

\section*{Acknowledgements}

LC was supported by the  NNSF of China (Grant No. 11871089), Beijing Natural Science Foundation (4173076), and the Fundamental Research Funds for the Central Universities (Grant Nos. KG12040501, ZG216S1810 and ZG226S18C1).  
The second author was supported in part by the National Sciences and Engineering Research Council (NSERC) of Canada Discovery 
Grant 5285.

\newpage

\appendix

\section{Representatives of equivalence classes of UOMs of four qubits}
\label{app1}

\

\centerline{\large{1. Representatives of $\UOM[8,4]$}}
\label{tab:8,4}
\

$
\noindent \text{Level 13} \\
\begin{array}{rlllllllll}
1 & addd & aDcc & acDC & aCCD & Abbb & AaBa & ABaA & AAAB & [1,4,4,4] \\
\end{array}
$

\

$
\noindent \text{Level 12} \\
\begin{array}{rlllllllll}
1 & acdd & aaDc & aAcd & aCCC & Abbb & AaaB & AABa & ABAA & [1,3,4,4] \\
\end{array}
$

\

$
\noindent \text{Level 11} \\
\begin{array}{rlllllllll}
1 & abdd & aaDc & aAcD & aBCC & Abbb & AaBa & AAaB & ABAA & [1,2,4,4] \\
2 & abdd & aaDc & aAcD & aBCC & Abbb & AaaB & AABa & ABAA & [1,2,4,4] \\
3 & acad & aacD & aAAc & aCCC & AbAb & Aaaa & AAbB & ABAA & [1,3,3,4] \\
4 & acad & aaAc & aAcD & aCCC & Abbb & AaaB & AABa & ABAA & [1,3,3,4] \\
5 & acad & aaAc & aAcD & aCCC & Abbb & AaAB & AABa & ABaA & [1,3,3,4] \\
6 & acca & aCac & aaCC & bAAA & Abba & ABab & AaBB & BAAA & [2,3,3,3] \\
\end{array}
$

\

$
\noindent \text{Level 10} \\
\begin{array}{rlllllllll}
1 & abad & aacD & aAAc & aBCC & Abbb & AaaB & AABa & ABAA & [1, 2, 3, 4] \\
2 & abad & aacD & aAAc & aBCC & AbAb & AabB & AAaa & ABBA & [1, 2, 3, 4] \\
3 & abad & aaAc & aAcD & aBCC & Abbb & AaBa & AAAB & ABaA & [1, 2, 3, 4] \\
4 & abad & aaAc & aAcD & aBCC & Abab & AabB & AAAa & ABBA & [1, 2, 3, 4] \\
5 & abad & aaAc & aAcD & aBCC & AbAb & AabB & AAaa & ABBA & [1, 2, 3, 4] \\
6 & abad & aaAc & aAcD & aBCC & Abab & AaAa & AAbB & ABBA & [1, 2, 3, 4] \\
7 & acca & aCac & aaAA & aACC & AAbb & AbaB & AaAA & ABBa & [1, 3, 3, 3] \\
8 & acca & aCac & aaAA & aACC & Aabb & AAaa & AbAB & ABBA & [1, 3, 3, 3] \\
\end{array}
$

\newpage

$
\noindent \text{Level 9} \\
\begin{array}{rlllllllll}
1 & abbd & aBac & aaAD & aABC & Abbb & ABaa & AaAB & AABA & [1, 2, 2, 4] \\
2 & abbd & aBac & aaAD & aABC & Abbb & ABaa & AAAB & AaBA & [1, 2, 2, 4] \\
3 & abbd & aaBc & aAaD & aBAC & AbBb & Aaba & AAAB & ABaA & [1, 2, 2, 4] \\
4 & abbd & aaBc & aAaD & aBAC & Abab & AaAa & AAbB & ABBA & [1, 2, 2, 4] \\
5 & abbd & aaBc & aAaD & aBAC & Abab & AaBB & AAAa & ABbA & [1, 2, 2, 4] \\
6 & abbd & aaBc & aAaD & aBAC & Abab & AaAa & AABB & ABbA & [1, 2, 2, 4] \\
7 & abca & aaaA & aACc & aBAC & AbAb & AaaA & AAbB & ABBa & [1, 2, 3, 3] \\
8 & abaa & aacA & aAAc & aBCC & Abbb & AaBa & AAaB & ABAA & [1, 2, 3, 3] \\
\end{array}
$

\newpage

$
\noindent \text{Level 8} \\
\begin{array}{rlllllllll}
1 & abba & aBac & aaAA & aABC & Abbb & ABaA & AaAB & AABa & [1, 2, 2, 3] \\
2 & abba & aBac & aaAA & aABC & Abba & ABab & AaAA & AABB & [1, 2, 2, 3] \\
3 & abba & aBac & aaAA & aABC & Abbb & AAaB & ABAa & AaBA & [1, 2, 2, 3] \\
4 & abba & aBac & aaAA & aABC & AbbA & AAaa & ABAb & AaBB & [1, 2, 2, 3] \\
5 & abbc & aaaC & aBAa & aABA & Abbb & ABaa & AaAB & AABA & [1, 2, 2, 3] \\
6 & abbc & aaaC & aBAa & aABA & Abbb & AaaB & ABAA & AABa & [1, 2, 2, 3] \\
7 & abbc & aaaC & aBAa & aABA & Abbb & AAaB & ABAA & AaBa & [1, 2, 2, 3] \\
8 & abba & aaBc & aAaA & aBAC & AbBb & Aaba & AAaB & ABAA & [1, 2, 2, 3] \\
\end{array}
$

\newpage

$
\noindent \text{Level 7} \\
\begin{array}{rlllllllll}
1 & abbb & aBaa & aaBA & aAAB & Abbb & ABaa & AaBA & AAAB & [1, 2, 2, 2] \\
2 & abbb & aBaa & aaAB & aABA & AbbB & ABaa & AaAb & AABA & [1, 2, 2, 2] \\
3 & abbb & aBaa & aaAB & aABA & AbbB & ABaA & AaAb & AABa & [1, 2, 2, 2] \\
4 & abbb & aBaa & aaAB & aABA & AbbB & Aaab & ABAA & AABa & [1, 2, 2, 2] \\
5 & abbb & aBaa & aaAB & aABA & AbbB & AAab & ABAa & AaBA & [1, 2, 2, 2] \\
6 & abbb & aaBa & aAaB & aBAA & AbBB & Aaba & AAab & ABAA & [1, 2, 2, 2] \\
7 & abbb & aaBa & aAaB & aBAA & AbBB & AabA & AAAb & ABaa & [1, 2, 2, 2] \\
8 & abbb & aaBa & aAaB & aBAA & AbBA & AabB & AAAa & ABab & [1, 2, 2, 2] \\
9 & abbb & aaBa & aAaB & aBAA & AbaB & Aabb & AAAA & ABBa & [1, 2, 2, 2] \\
10 & abbb & aaBa & aAaB & aBAA & AbAB & Aabb & AAaa & ABBA & [1, 2, 2, 2] \\
11 & abbb & aaBa & aAaB & aBAA & AbAB & Aabb & AAaA & ABBa & [1, 2, 2, 2] \\
12 & abbb & aaaB & aABa & aBAA & AbBB & Aaba & AAab & ABAA & [1, 2, 2, 2] \\
13 & abbb & aaaB & aABa & aBAA & AbBB & AabA & AAab & ABAa & [1, 2, 2, 2] \\
14 & abbb & aaaB & aABa & aBAA & AbBB & AabA & AAAb & ABaa & [1, 2, 2, 2] \\
15 & abbb & aaaB & aABa & aBAA & Abaa & AabA & AAAb & ABBB & [1, 2, 2, 2] \\
16 & abbb & aaBa & aAaB & aBAA & AbaB & AaAa & AAbb & ABBA & [1, 2, 2, 2] \\
17 & abbb & aaBa & aAaB & aBAA & AbaB & AaAA & AAbb & ABBa & [1, 2, 2, 2] \\
18 & abbb & aaBa & aAaB & aBAA & AbAB & AaaA & AAbb & ABBa & [1, 2, 2, 2] \\
\end{array}
$

\

\

\centerline{\large{2. Representatives of $\UOM[9,4]$}}

\

$
\noindent \text{Level 10} \\
\begin{array}{rllllllllll}
1 & aaca & aaCa & abBA & bAaa & bBAA & Aaac & AaaC & AbAB & BAbb & [2, 2, 3, 3] \\
\end{array}
$

\

$
\noindent \text{Level 9} \\
\begin{array}{rllllllllll}
1 & aaac & aaaC & aaAb & aAaA & bAba & Aaaa & AABb & AbbA & BBAB & [2, 2, 2, 3] \\
2 & aaAa & aAbb & baac & baaC & bAAB & AAaa & AbAb & Baaa & BBBA & [2, 2, 2, 3] \\
3 & aaac & aaaC & aAbB & baAa & bABA & Abaa & ABaa & AbbA & BBAb & [2, 2, 2, 3] \\
4 & aaac & aaaC & aAbb & baAa & bABA & Aaaa & AAaa & AbbA & BBAB & [2, 2, 2, 3] \\
\end{array}
$

\

$
\noindent \text{Level 8} \\
\begin{array}{rllllllllll}
1 & aaaa & aAaa & abAb & baaA & bAAB & Aaaa & AaAa & AAbb & BBBA & [2, 2, 2, 2] \\
2 & abaa & aBaa & aBAb & baaA & bAAB & Aaba & AaBa & AAbb & BbBA & [2, 2, 2, 2] \\
3 & aaAa & aAbb & baaa & baaA & bAAB & AAaa & AbAb & Baaa & BBBA & [2, 2, 2, 2] \\
4 & abAa & aabA & baaa & bAaa & bAAA & AbaA & AaAb & Bbaa & BBBB & [2, 2, 2, 2] \\
5 & aaaa & aaAa & aAbb & baaA & bAAB & Abaa & ABaa & ABAb & BbBA & [2, 2, 2, 2] \\
6 & abAa & aabA & aAab & baaa & bAAA & AAba & AbaA & AaAb & BBBB & [2, 2, 2, 2] \\
\end{array}
$

\

\

\centerline{\large{3. Representatives of $\UOM[10,4]$}}

\

$
\noindent \text{Level 11} \\
\begin{array}{rlllllllllll}
1 & aabd & aabD & aaBc & aaBC & aAAb & bAaa & Aaaa & AbAb & ABCA & BAcB &  [2, 2, 3, 4] \\
2 & aaad & aaaD & aaAc & aaAC & aAcA & bAba & Aaba & AbBB & ABcA & BACb &  [2, 2, 3, 4] \\
\end{array}
$

\newpage

$
\noindent \text{Level 10} \\
\begin{array}{rlllllllllll}
1 & aaad & aaaD & aaAb & aAbb & bBAB & Aaac & AaaC & AAaa & AbAa & BABA & [2, 2, 2, 4] \\
2 & aaad & aaaD & aAaA & abAA & baAa & bABa & Aaac & AaaC & AAbb & BBAB & [2, 2, 2, 4] \\
3 & aaad & aaaD & abAA & baAa & bABa & Aaac & AaaC & AAbB & BAab & BBAb & [2, 2, 2, 4] \\
4 & aaac & aaAa & aaAA & aaaC & aAca & bAbA & AabA & Abca & ABBB & BACb & [2, 2, 3, 3] \\
5 & aabc & aaBa & aaBA & aabC & aAAa & bAcA & Aaab & AbAa & ABCB & BAab & [2, 2, 3, 3] \\
6 & aaba & aaBa & aacA & aaCA & aAAc & bAab & Aaab & AbAc & ABbB & BABC & [2, 2, 3, 3] \\
7 & aaba & aaBa & aabA & aaBA & aAAc & bAab & Aaab & AbAc & ABCB & BAcC & [2, 2, 3, 3] \\
8 & aaca & aaCa & aAba & abbA & bABb & Aaac & AaAa & AaaC & AAaB & BBAA & [2, 2, 3, 3] \\
9 & aaac & aaAa & aaaC & aAab & bAAa & bbAA & Aaca & AaCa & ABBA & BAbB & [2, 2, 3, 3] \\
10 & abbA & aaca & aaCa & aAba & bABB & AbAb & Aaac & AaaC & AAab & BBAA & [2, 2, 3, 3] \\
11 & aAac & aAaC & aAAb & bacA & baba & baBa & AAaa & ABCA & Baaa & BbAB & [2, 2, 3, 3] \\
12 & aaAc & aaAC & aAAb & bbad & bbaD & bBaa & AaAa & ABBA & Bbab & BAbB & [2, 2, 2, 4] \\
13 & aBab & aaAc & aaAC & bbad & bbaD & bAAa & AaAa & ABBA & Bbab & BAbB & [2, 2, 2, 4] \\
14 & aaAb & aaAB & bbad & bbaD & bBaa & bAAa & AaAa & ABBA & Bbac & BAbC & [2, 2, 2, 4] \\
15 & aAab & aAaB & baca & baCa & babA & bAAa & AAaa & ABBA & Babc & BbAC & [2, 2, 3, 3] \\
16 & aAac & aAaC & aAAb & baca & babA & baCa & AAbb & AbBA & Baaa & BBAB & [2, 2, 3, 3] \\
17 & abaa & aaAa & aBac & aBaC & baAA & bAAB & Aaca & AaCa & AAbb & BbBA & [2, 2, 3, 3] \\
\end{array}
$

\newpage

$
\noindent \text{Level 9} \\
\begin{array}{rlllllllllll}
1 & aaaa & aaaA & aaAb & aAbb & bBAB & Aaac & AaaC & AAaa & AbAa & BABA & [2, 2, 2, 3] \\
2 & aaaa & aaaA & aaAa & aAba & bBAA & Aaac & AaaC & AAab & AbAb & BABB & [2, 2, 2, 3] \\
3 & aaaa & aaaA & aaAc & aAbc & bBAC & Aaaa & AaaA & AAab & AbAb & BABB & [2, 2, 2, 3] \\
4 & aaac & aaaC & aAaa & abAa & baAA & bABA & Aaaa & AaaA & AAbb & BBAB & [2, 2, 2, 3] \\
5 & aaac & aaaC & aAaA & abAA & baAa & bABa & Aaab & AaaB & AAbB & BBAb & [2, 2, 2, 3] \\
6 & aaaa & aaaA & aAaa & abAa & baAA & bABA & Aaac & AaaC & AAbb & BBAB & [2, 2, 2, 3] \\
7 & aaab & aaaB & aAaA & abAA & baAa & bABa & Aaac & AaaC & AAbB & BBAb & [2, 2, 2, 3] \\
8 & aaab & aaaB & aAaA & abAA & baAa & bABa & Aaab & AaaB & AAbc & BBAC & [2, 2, 2, 3] \\
\end{array}
$

\newpage

$
\noindent \text{Level 8} \\
\begin{array}{rlllllllllll}
1 & aaaa & aAaa & aaAa & aabA & bAAB & Aaaa & AAaa & AbaA & AbAb & BBBA & [2, 2, 2, 2] \\
2 & aAaa & aAaA & aAAb & baaa & baAa & babA & AAaa & ABBA & Baaa & BbAB & [2, 2, 2, 2] \\
3 & aAaa & aaAa & aabA & baaa & bAaA & bAAb & AAaa & AbAB & Baaa & BBBA & [2, 2, 2, 2] \\
4 & aAaa & aaaA & aaAb & baaa & bBAB & AAaa & AbAa & AbBA & Baaa & BAbA & [2, 2, 2, 2] \\
5 & aaaa & aaaA & aaAb & bAaa & bBAB & Aaaa & AbAa & AbBA & BAaa & BAbA & [2, 2, 2, 2] \\
6 & aaaa & aAaa & aBAB & bbAa & bbBA & Aaaa & AAaa & AAbA & BaaA & BaAb & [2, 2, 2, 2] \\
7 & abaa & aBaa & aAbA & bbAa & bbBA & Aaaa & AAaa & ABAB & BaaA & BaAb & [2, 2, 2, 2] \\
8 & abaa & aBaa & aBAB & bbAa & bbBA & Aaaa & AAaa & AAbA & BaaA & BaAb & [2, 2, 2, 2] \\
\end{array}
$

\

\

\centerline{\large{4. Representatives of irreducible classes in $\UOM[12,4]$}}

\

In $\cO(12,4)$ there are 1209 equivalence classes of UOMs of which only 161 are irreducible. While all the equivalence classes occupy the levels 7-14, the irreducible ones occupy 
only the levels 8-12. We list below the representatives of the 161 irreducible classes.

\

$
\noindent \text{Level 12} \\
\begin{array}{rlllllllllllll}
1 & aaaa & aaAe & aaAE & abaA & aBaA & bAAb & Aabd & AabD & AaBc & AaBC & AAcB & BACa & [2, 2, 3, 5] \\
2 & aaba & aadA & aaDA & abBa & aBBa & bAbb & Aaad & AaAc & AaAC & AaaD & AAcB & BACA & [2, 2, 4, 4] \\
\end{array}
$

\newpage

$
\noindent \text{Level 11} \\
\begin{array}{rlllllllllllll}
1 & aAaa & aAcA & aACA & baab & babB & baBB & AAac & AAaC & AbAb & Baad & BaaD & BBAa & [2, 2, 3, 4] \\
2 & acaa & abAa & aCaa & aBAb & aBAB & bAaA & Aaba & AAca & AACa & AaBc & AaBC & BbbA & [2, 3, 3, 3] \\
3 & aaaa & aAae & aAaE & aabA & aaBA & bAAb & Aaad & AaaD & AAac & AAaC & AbAB & BBAa & [2, 2, 2, 5] \\
4 & aAad & aAaD & aAAb & baaa & babA & baBA & AAac & AAaC & AbAa & Baae & BaaE & BBAB & [2, 2, 2, 5] \\
5 & aaca & aaaA & aaCa & abAA & aBAA & bAab & Aaad & AaaD & AaAc & AaAC & AAbB & BABa & [2, 2, 3, 4] \\
6 & aaca & aabA & aaCa & abBA & aBBA & bAAa & Aaad & AaaD & AaAc & AaAC & AAab & BAbB & [2, 2, 3, 4] \\
7 & aaab & aaAd & aaAD & abaB & aBaB & bAAc & Aaca & AaaA & AaAA & AaCa & AAbC & BABb & [2, 2, 3, 4] \\
8 & aaab & aaAd & aaAD & abaB & aBaB & bAAc & Aaca & AabA & AaBA & AaCa & AABC & BAbb & [2, 2, 3, 4] \\
9 & aaba & aaaA & aaAA & abBa & aBBa & bAbb & Aaad & AaaD & AaAc & AaAC & AAcB & BACA & [2, 2, 3, 4] \\
10 & aaaa & aaAd & aaAD & abaA & aBaA & bAAb & Aaba & AabA & AaBc & AaBC & AAcB & BACa & [2, 2, 3, 4] \\
11 & aaab & aaAd & aaAD & abaB & aBaB & bAAA & Aaba & AabA & AaBc & AaBC & AAca & BACb & [2, 2, 3, 4] \\
12 & aaab & aaAd & aaAD & abaB & aBaB & bAAc & Aaba & AabA & AaBa & AaBA & AAcC & BACb & [2, 2, 3, 4] \\
13 & aaaa & aaAa & aaAA & abaA & aBaA & bAAb & Aabd & AabD & AaBc & AaBC & AAcB & BACa & [2, 2, 3, 4] \\
14 & aaab & aaAa & aaAA & abaB & aBaB & bAAA & Aabd & AabD & AaBc & AaBC & AAca & BACb & [2, 2, 3, 4] \\
15 & aaab & aaAa & aaAA & abaB & aBaB & bAAc & Aaba & AabA & AaBd & AaBD & AAcC & BACb & [2, 2, 3, 4] \\
16 & aaca & aabA & aaBA & abCa & aBCa & bABA & Aaad & AaAc & AaAC & AaaD & AAbb & BAcB & [2, 2, 3, 4] \\
17 & acaa & aacA & aaCA & aAaA & aAAc & aAAC & aCaa & baAa & Abab & AAbB & ABBA & BaAa & [2, 3, 3, 3] \\
18 & aaac & aAaa & abAa & aAcA & aaAA & aACA & aaaC & bBAa & Acab & ACbA & AbBB & BBAa & [2, 3, 3, 3] \\
\end{array}
$

\newpage

$
\noindent \text{Level 10} \\
\begin{array}{rlllllllllllll}
1 & abaB & aaAA & bbab & bBad & bBaD & baAa & bAAc & bAAC & AbaB & AaAA & BAbb & BBBa & [2, 2, 2, 4] \\
2 & aaBA & aAaB & baca & babA & baCa & bAab & bAAc & bAAC & AaBA & AAaB & BbAa & BBbb & [2, 2, 3, 3] \\
3 & aaba & aAAa & aaBb & aaBB & bAaa & bBbA & Aaca & AaCa & AAAa & AbAA & BAac & BAaC & [2, 2, 3, 3] \\
4 & aAAa & aAac & aAaC & baba & baBb & baBB & AAca & AACa & AbbA & Baaa & BaAa & BBAA & [2, 2, 3, 3] \\
5 & aAba & aABc & aABC & baaa & baAb & baAB & AAca & AACa & AbaA & Baaa & BaAa & BBbA & [2, 2, 3, 3] \\
6 & aAaa & aAAc & aAAC & baca & baCb & baCB & AAba & AABa & AbcA & Baba & BaBa & BBaA & [2, 2, 3, 3] \\
7 & aaaa & aAad & aAaD & aabA & aaBA & bAAb & Aaac & AaaC & AAaa & AAaA & AbAB & BBAa & [2, 2, 2, 4] \\
8 & aaab & aAad & aAaD & aabB & aaBB & bAAA & Aaac & AaaC & AAaa & AAaA & AbAa & BBAb & [2, 2, 2, 4] \\
\end{array}
$

$
\begin{array}{rlllllllllllll}
43 & aaab & aaAa & aaAA & abaB & aBaB & bAcb & Aaba & AaBa & AabA & AaBA & AACc & BAAC & [2, 2, 3, 3] \\
44 & aaaa & aaAa & aaAA & abaA & aBaA & bAca & Aabb & AaBb & AabB & AaBB & AACc & BAAC & [2, 2, 3, 3] \\
45 & aaab & aaAa & aaaB & abAA & aBAA & bAca & Aabb & AaBa & AaBA & AabB & AACc & BAaC & [2, 2, 3, 3] \\
46 & aacA & aaCA & abAa & aBAa & baaa & bAab & AaaA & AaAc & AaAC & AAbB & Baaa & BABA & [2, 2, 3, 3] \\
47 & aaAa & aabA & abBA & aBBA & baaa & bAca & AaAc & AaaA & AaAC & AACb & Baaa & BAbB & [2, 2, 3, 3] \\
48 & aaAa & aacA & aaCA & aABA & baaa & bAbb & AaAc & AaaA & AaAC & AAAB & Bbaa & BBaa & [2, 2, 3, 3] \\
49 & aaAa & aacA & aaCA & aAAB & baaa & bAbb & AaAc & AaaA & AaAC & AABA & Bbaa & BBaa & [2, 2, 3, 3] \\
50 & aAad & abAd & aBAb & aBAB & baad & bbbD & AAbd & AaAd & AABc & AABC & Baaa & BaaA & [2, 2, 2, 4] \\
51 & abAa & aAaa & aBAc & aBAC & baaa & bbbA & AaAa & AAca & AACa & AABA & Baab & BaaB & [2, 2, 3, 3] \\
\end{array}
$

\newpage

$
\noindent \text{Level 9} \\
\begin{array}{rlllllllllllll}
1 & aaaB & aBAa & baab & bAac & bAaC & bbAa & baAA & bAAA & AaaB & ABAa & BAbA & BbBb & [2, 2, 2, 3] \\
2 & abaB & aaAa & bbab & bBaa & bBaA & baAA & bAAc & bAAC & AbaB & AaAa & BAbb & BBBA & [2, 2, 2, 3] \\
3 & abaA & aaAB & bbaa & bBaa & bBaA & baAb & bAAc & bAAC & AbaA & AaAB & BAba & BBBb & [2, 2, 2, 3] \\
4 & abaC & aaAA & bbac & bBab & bBaB & baAa & bAAb & bAAB & AbaC & AaAA & BAbc & BBBa & [2, 2, 2, 3] \\
5 & aaaa & aaAa & aAAa & aabA & bAaa & bbBA & Aaaa & AbAa & ABAc & ABAC & BAab & BAaB & [2, 2, 2, 3] \\
6 & abAa & aBAc & aBAC & baaa & bAaa & bbBA & AaAa & AAAa & AAbA & BAaa & Baab & BaaB & [2, 2, 2, 3] \\
7 & aaAa & aAAc & aAAC & bbaa & bBaa & baBA & AaAa & AAAa & ABbA & BBaa & Bbab & BbaB & [2, 2, 2, 3] \\
8 & aBaa & aAAc & aAAC & bbaa & baAa & baBA & ABaa & AAAa & ABbA & BaAa & Bbab & BbaB & [2, 2, 2, 3] \\
\end{array}
$

$
\begin{array}{rlllllllllllll}
46 & aAaa & aAaA & aAAA & baab & babB & baBB & AAac & AAaC & AbAb & Baaa & BaaA & BBAa & [2, 2, 2, 3] \\
47 & aAaa & aAaA & aAAB & baaa & babA & baBA & AAac & AAaC & AbAa & Baab & BaaB & BBAb & [2, 2, 2, 3] \\
48 & aAab & aAaB & aAAB & baaa & babA & baBA & AAac & AAaC & AbAa & Baaa & BaaA & BBAb & [2, 2, 2, 3] \\
49 & aAaa & aAaA & aAAc & baab & babB & baBB & AAaa & AAaA & AbAb & Baaa & BaaA & BBAC & [2, 2, 2, 3] \\
50 & aAab & aAaB & aAAc & baaa & babA & baBA & AAaa & AAaA & AbAa & Baab & BaaB & BBAC & [2, 2, 2, 3] \\
51 & aAaa & aAaA & aAAc & baaa & babA & baBA & AAab & AAaB & AbAa & Baab & BaaB & BBAC & [2, 2, 2, 3] \\
52 & aAab & aAaB & aAAc & baaa & babA & baBA & AAab & AAaB & AbAa & Baaa & BaaA & BBAC & [2, 2, 2, 3] \\
53 & aAab & aAaB & aAAA & baac & babC & baBC & AAaa & AAaA & AbAc & Baab & BaaB & BBAa & [2, 2, 2, 3] \\
\end{array}
$

$
\noindent \text{Level 8} \\
\begin{array}{rlllllllllllll}
1 & aaBA & aAAB & aBaa & bbaa & baAa & babA & bAAb & bAaA & AaBA & AAAB & ABaa & Bbbb & [2, 2, 2, 2] \\
2 & aABa & aBAA & baaa & baAa & bAba & bbAA & baaA & bAaA & AABa & ABAA & Baab & BbbB & [2, 2, 2, 2] \\
3 & aBaa & aaBA & bbaa & bbAa & bBAa & babA & bAaA & bAAA & ABaa & AaBA & BAAb & BbbB & [2, 2, 2, 2] \\
4 & aBba & aaAA & bbba & bbBa & bBBa & baaA & bAaA & bAAA & ABba & AaAA & BABb & BbaB & [2, 2, 2, 2] \\
5 & aBba & aaAA & bbba & baBa & bABa & bbaA & bAAA & bBaA & ABba & AaAA & BABb & BbaB & [2, 2, 2, 2] \\
6 & aaaa & aAaa & aAAa & aaAb & aaAB & bbaA & Aaaa & AaAa & AAaa & AAAa & ABBA & BAbA & [2, 2, 2, 2] \\
7 & aAaa & aaAa & aAAa & abAA & baaa & bBBA & AAaa & AaAa & AAAa & AAbA & Baab & BaaB & [2, 2, 2, 2] \\
8 & aaaa & aAaa & aAAa & aBBA & baAa & bbAA & AAaa & AAAa & Aaab & AaaB & BaAa & BAbA & [2, 2, 2, 2] \\
\end{array}
$

\section{Arrows of the Hasse diagrams of $\UOM[m,4]$} \label{app2}

For $\UOM[m,4]$, $m=8,9,10$, we list all the arrows from level $l$ to $l-1$, for all appropriate levels $l$. In the case $m=12$ we list only the arrows which join irreducible equivalence classes.

\

\centerline{5. The case $m=8$}

Arrows from level $11$ to $10$: 
$$
\begin{array}{ll}
1 \to 1,2,3,6      & d_4=a_4,d_4=B_4,d_4=A_4,d_4=b_4 \\
2 \to 4,5          & d_4=b_4,d_4=B_4 \\
3 \to 1,5,8,9,     & c_3=B_3,c_3=b_3,c_4=A_4,c_4=a_4, \\
3 \to 13,15,16     & d_4=a_4,d_4=b_4,d_4=B_4 \\
4 \to 2,3,4,5,9,   & c_2=B_2,c_3=b_3,c_3=B_3,c_2=b_2,c_4=b_4, \\
4 \to 10,12,13,14,15 
                   & c_4=B_4,d_4=b_4,c_4=A_4,d_4=A_4,d_4=a_4 \\
5 \to 4,6,7,10,    & c_3=b_3,c_3=B_3,c_4=B_4,c_4=b_4, \\
5 \to 11,13,14     & d_4=A_4,c_4=a_4,d_4=a_4 \\
6 \to 7,20,21      & b_1=a_1,c_4=b_4,c_4=B_4 \\
\end{array}
$$

Arrows from level $10$ to $9$: 
$$
\begin{array}{ll}
1 \to 4,6,11,12,14,18      & c_3=b_3,c_3=B_3,d_4=a_4,d_4=A_4,
d_4=b_4,d_4=B_4 \\
2 \to 2,3,17,18,19,20      & c_3=b_3,c_3=B_3,d_4=b_4,d_4=a_4,
d_4=A_4,d_4=B_4 \\
3 \to 3,6,12,13,15,19      & c_3=B_3,c_3=b_3,d_4=a_4,d_4=A_4,
d_4=b_4,d_4=B_4 \\
4 \to 2,5,7,9,10,21,22,24  & c_3=b_3,c_3=B_3,d_4=b_4,c_4=A_4,
c_4=a_4,d_4=A_4,d_4=B_4,d_4=a_4 \\
5 \to 5,6,8,9,22,23,24,25  & c_3=b_3,d_4=a_4,c_4=A_4,c_4=a_4,
d_4=b_4,d_4=a_4,d_4=A_4,d_4=B_4 \\
6 \to 1,2,14,15,16,17      & c_3=b_3,c_3=B_3,d_4=a_4,d_4=A_4,
d_4=b_4,d_4=B_4 \\
7 \to 7,16                 & c_2=b_2,c_4=B_4 \\
8 \to 8,11                 & c_2=B_2,c_2=b_2 \\
9 \to 9,12,20,25           & c_3=B_3,c_3=b_3,c_2=B_2,c_2=b_2 \\
10 \to 10,13,17,22         & c_3=B_3,c_3=b_3,c_2=B_2,c_2=b_2 \\
11 \to 10,16               & c_2=b_2,c_2=B_2 \\
12 \to 19,24               & c_2=b_2,c_2=B_2 \\
13 \to 14,15,18,21,23,24   & c_2=b_2,c_3=b_3,c_4=b_4,c_3=B_3,
c_4=B_4,c_2=B_2 \\
14 \to 7,9,13,17           & c_3=b_3,c_2=B_2,c_3=B_3,c_2=b_2 \\
15 \to 8,12,20,22          & c_2=B_2,c_3=B_3,c_2=b_2,c_3=b_3 \\
16 \to 11,25           & c_2=B_2,c_2=b_2 \\
17 \to 26              & c_4=A_4 \\
18 \to 27              & d_4=A_4 \\
19 \to 27,28           & c_4=A_4,c_4=b_4 \\
20 \to 16,30,31        & b_2=a_2,c_3=b_3,c_3=B_3 \\
21 \to 7,31,33         & b_2=a_2,c_3=B_3,c_3=b_3 \\
22 \to 34,35,36        & b_4=A_4,d_4=c_4,c_4=a_4 \\
23 \to 29,34,35        & c_3=a_3,c_4=B_4,c_3=b_3 \\
24 \to 37              & b_4=a_4  \\
25 \to 30,37,39,40     & b_3=a_3,c_4=b_4,c_4=B_4,c_3=b_3 \\
26 \to 32,41,43        & b_4=a_4,c_4=b_4,c_4=B_4 \\
27 \to 38,39,41        & c_4=a_4,d_4=C_4,b_4=a_4 \\
28 \to 1,40,42,43      & b_1=a_1,b_4=a_4,c_4=a_4,b_4=A_4 \\
29 \to 3,44,45         & b_1=a_1,b_4=a_4,b_4=A_4 \\
\end{array}
$$

Arrows from level $9$ to $8$: 
$$
\begin{array}{ll}
1 \to 1,2,6        & d_4=a_4,c_4=a_4,c_4=A_4 \\
2 \to 5,7,8,9      & c_4=a_4,c_4=A_4,d_4=a_4,d_4=A_4 \\
3 \to 10,21,22,23  & d_4=A_4,c_4=A_4,c_4=a_4,d_4=a_4 \\
4 \to 12,19,20     & c_4=a_4,d_4=a_4,d_4=A_4 \\
5 \to 3,4,11,15,16 & c_4=a_4,c_4=b_4,c_4=B_4,d_4=A_4,d_4=a_4 \\
6 \to 13,14,17,18  & c_4=a_4,c_4=A_4,d_4=A_4,d_4=a_4 \\
7 \to 5,11         & c_3=B_3,c_3=b_3 \\
8 \to 4,18         & c_3=B_3,c_3=b_3 \\
\end{array}
$$

Arrows from level $8$ to $7$: 
$$
\begin{array}{llcll}
1 \to 9,17   & c_4=b_4,c_4=B_4 & & 22 \to 3,4 &c_4=B_4,c_4=b_4\\
2 \to 1,2    & c_4=b_4,c_4=B_4 & & 23 \to 17,18 & c_4=b_4,
                                                  c_4=B_4 \\
3 \to 10,11  & c_4=B_4,c_4=b_4 &\quad& 24 \to 5    & b_4=A_4 \\   
4 \to 5,13   & c_4=b_4,c_4=B_4 & & 26 \to 1    & b_4=a_4 \\
5 \to 2,6    & c_4=b_4,c_4=B_4 & & 27 \to 2    & b_4=a_4 \\
6 \to 2,3    & c_4=b_4,c_4=B_4 & & 28 \to 6    & b_4=a_4 \\
7 \to 4,6    & c_4=B_4,c_4=b_4 & & 29 \to 3    & b_4=A_4 \\
8 \to 11,16  & c_4=B_4,c_4=b_4 & & 30 \to 10   & b_4=A_4 \\
9 \to 11,18  & c_4=b_4,c_4=B_4 & & 31 \to 8    & b_4=A_4 \\
10 \to 8,18  & c_4=B_4,c_4=b_4 & & 32 \to 12   & b_4=a_4 \\
11 \to 6,12  & c_4=B_4,c_4=b_4 & & 33 \to 2    & b_4=A_4 \\
12 \to 16,17 & c_4=B_4,c_4=b_4 & & 35 \to 6    & b_4=A_4 \\
13 \to 9,11  & c_4=b_4,c_4=B_4 & & 36 \to 9    & b_4=A_4 \\              
14 \to 11,18 & c_4=b_4,c_4=B_4 & & 37 \to 1    & b_1=a_1 \\
15 \to 5,6   & c_4=B_4,c_4=b_4 & & 38 \to 2    & b_1=a_1 \\
16 \to 12,13 & c_4=b_4,c_4=B_4 & & 39 \to 3    & b_4=a_4 \\
17 \to 4,13  & c_4=b_4,c_4=B_4 & & 40 \to 7,8  & b_4=A_4,
                                                 b_1=A_1 \\
18 \to 13,14 & c_4=b_4,c_4=B_4 & & 41 \to 2    & b_1=a_1 \\
19 \to 14,15 & c_4=B_4,c_4=b_4 & & 42 \to 3,17 & b_1=a_1,
                                                 b_2=a_2 \\
20 \to 7,14  & c_4=b_4,c_4=B_4 & & 43 \to 4,18 & b_3=A_3,
                                                 b_4=A_4 \\
21 \to 4,7   & c_4=b_4,c_4=B_4 & &             & \\
\end{array}
$$

\

\centerline{6. The case $m=9$}

The UOMs occupy the levels $8-10$. There are only two arrows 
from level 10 to 9: 
$$
1\to 3,4 \quad c_3=b_3,c_3=a_3
$$
and eight arrows from level 9 to 8:

$$
\begin{array}{llcll}
1 \to 1,3 & c_4=a_4,c_4=b_4 & \quad & 3 \to 2,5 & c_4=b_4,
                                                  c_4=a_4 \\
2 \to 3,4 & c_4=a_4,c_4=b_4 & & 4 \to 1,5 & c_4=a_4,c_4=b_4 \\
\end{array}
$$

\

\centerline{7. The case $m=10$}

Arrows from level 11 to 10:
$$
\begin{array}{ll}
1 \to 1,3,5,7 & b_3=a_3,c_3=B_3,c_4=a_4,d_4=c_4 \\
2 \to 2,4,6,7 & b_3=a_3,c_4=a_4,c_4=b_4,d_4=c_4 \\
\end{array}
$$

Arrows from level 10 to 9:
$$
\begin{array}{ll}
1 \to 1,2,3           & d_4=a_4,d_4=b_4,d_4=c_4 \\
2 \to 4,5,6,7,8       &c_4=a_4,c_4=b_4,d_4=a_4,d_4=b_4,d_4=c_4\\
3 \to 11,12,13        & d_4=a_4,d_4=b_4,d_4=c_4 \\
4 \to 3,4,6,9         & c_4=a_4,b_3=a_3,b_3=A_3,c_4=b_4 \\
5 \to 1,2,8,10,11,12  & b_3=A_3,b_3=a_3,c_4=a_4,c_4=b_4,c_3=B_3,
                        c_3=b_3 \\
6 \to 5,7,9,13        & b_4=A_4,b_4=a_4,c_3=a_3,c_3=b_3 \\
7 \to 3,8,13          & b_3=a_3,b_4=a_4,c_3=B_3 \\
8 \to 2,14,15,16      & c_4=a_4,c_3=a_3,c_3=b_3,c_4=b_4 \\
9 \to 6,14,18,19      & c_4=a_4,c_3=A_3,c_4=b_4,c_3=b_3 \\
10 \to 20,21          & c_4=a_4,c_3=b_3 \\
11 \to 15,25          & b_3=a_3,c_3=b_3 \\
12 \to 27,28          & d_4=b_4,d_4=c_4 \\
13 \to 29,30,31       & d_4=a_4,d_4=b_4,d_4=c_4 \\
14 \to 25,26,32,33,34 &b_4=a_4,d_4=a_4,c_4=b_4,d_4=b_4,d_4=c_4\\
15 \to 16,22,26,35    & b_4=a_4,c_3=a_3,c_3=b_3,c_4=b_4 \\
16 \to 21,23,27,30    & c_3=a_3,c_4=a_4,c_4=b_4,c_3=b_3 \\
17 \to 18,22,36,37    & c_4=a_4,c_3=a_3,c_4=b_4,c_3=B_3 \\
\end{array}
$$

Arrows from level 9 to 8:
$$
\begin{array}{llll}
1 \to 1,3      & c_4=a_4,c_4=b_4 & 
20 \to 13,14   & c_4=a_4,c_4=b_4 \\
2 \to 1,2      & c_4=a_4,c_4=b_4 & 
21 \to 14,16   & c_4=a_4,c_4=b_4 \\
3 \to 1        & c_4=a_4 & 
22 \to 9,17    & c_4=a_4,c_4=b_4 \\
4 \to 1,4      & c_4=a_4,c_4=b_4 & 
23 \to 14,18   & c_4=a_4,c_4=b_4  \\
5 \to 5,6      & c_4=a_4,c_4=b_4 & 
24 \to 15,19   & c_4=a_4,c_4=b_4 \\
6 \to 1,5      & c_4=a_4,c_4=b_4 & 
25 \to 10,20   & c_4=a_4,c_4=b_4 \\
7 \to 4,6      & c_4=a_4,c_4=b_4 & 
26 \to 10,21   & b_4=a_4,c_4=b_4 \\
8 \to 1,6      & b_4=a_4,c_4=b_4 & 
27 \to 16,22   & c_4=a_4,c_4=b_4 \\
9 \to 4,5      & c_4=a_4,c_4=A_4 & 
28 \to 16      & b_4=a_4 \\
10 \to 2,3,7,8 & b_4=A_4,b_4=a_4,c_4=B_4,c_4=b_4 & 
29 \to 18,23   & c_4=a_4,c_4=b_4 \\
11 \to 6,8     & c_4=a_4,c_4=b_4 & 
30 \to 18,22   & c_4=a_4,b_4=a_4\\
12 \to 6,7     & c_4=a_4,c_4=b_4 & 
31 \to 18      & c_4=a_4 \\
13 \to 6       & c_4=a_4         & 
32 \to 19,21   & c_4=a_4,c_4=b_4 \\
14 \to 1,9     & c_4=a_4,c_4=b_4 & 
33 \to 10,19   & b_4=a_4,c_4=A_4 \\
15 \to 2,10    & c_4=a_4,c_4=b_4 & 
34 \to 19,20   & c_4=a_4,c_4=b_4 \\
16 \to 9,10    & c_4=a_4,c_4=b_4 & 
35 \to 17,21   & c_4=a_4,c_4=b_4 \\
17 \to 11,13   & c_4=b_4,c_4=a_4 & 
36 \to 17,24   & c_4=a_4,c_4=b_4 \\
18 \to 9,12    & c_4=b_4,c_4=a_4 & 
37 \to 12,24   & c_4=a_4,c_4=b_4 \\
19 \to 5,12    & c_4=a_4,c_4=b_4 & 
               & \\
\end{array}
$$

\

\centerline{8. The case $m=12$}

Arrows from level 12 to 11:
$$
\begin{array}{ll}
1 \to 3,4,10,11    & b_3=a_3,c_3=b_3,d_4=a_4,d_4=b_4, \\
1 \to 12,13,14,15  & d_4=c_4,e_4=a_4,e_4=b_4,e_4=d_4 \\
2 \to 5,6,7,8,     & b_3=a_3,c_3=a_3,c_4=a_4,c_4=b_4, \\
2 \to 9,12,13,16   & d_3=a_3,d_4=c_4,d_3=b_3,d_3=c_3 \\
\end{array}
$$

Arrows from level 11 to 10:
$$
\begin{array}{ll}
1 \to 1,4,5,6,19      &c_3=b_3,d_4=A_4,d_4=b_4,d_4=c_4,c_3=a_3\\
2 \to 2,3,5,31        & c_4=b_4,c_2=a_2,c_2=b_2,c_4=a_4 \\
3 \to 7,8,9,          & c_4=a_4,c_4=b_4,e_4=a_4, \\
3 \to 10,11,12        & e_4=b_4,e_4=c_4,e_4=d_4 \\
4 \to 16,17,18,19,20,&c_4=a_4,c_4=b_4,d_4=a_4,d_4=b_4,d_4=c_4,\\
4 \to 21,22,23,24     & e_4=a_4,e_4=b_4,e_4=c_4,e_4=d_4 \\
5 \to 9,13,25,26,     & c_3=a_3,c_3=b_3,c_4=a_4,c_4=b_4,\\
5 \to 27,28,29        & d_4=a_4,d_4=b_4,d_4=c_4,\\
6 \to 14,21,30,31,32  &c_3=a_3,c_3=b_3,c_4=b_4,d_4=b_4,d_4=c_4\\
7 \to 11,15,25,34,    & c_3=a_3,c_3=b_3,b_3=a_3,c_4=a_4, \\
7 \to 35,36,37,38     & c_4=A_4,d_4=a_4,d_4=b_4,d_4=c_4 \\
8 \to 15,20,26,30,    & b_3=a_3,c_3=b_3,b_4=a_4,c_4=A_4, \\
8 \to 31,41,42        & c_4=a_4,d_4=a_4,d_4=b_4 \\
9 \to 9,14,36,41,43  & b_3=A_3,c_3=a_3,d_4=a_4,d_4=b_4,d_4=c_4\\
10 \to 7,9,16,18,     & b_3=A_3,b_3=a_3,c_3=B_3,c_3=b_3, \\
10 \to 33,39,45       & c_4=b_4,d_4=b_4,d_4=c_4\\
11 \to 8,10,17,19,    & b_3=A_3,b_3=a_3,c_3=B_3,c_3=b_3, \\
11 \to 32,33,40,42    & c_4=a_4,c_4=b_4,d_4=a_4,d_4=b_4 \\
12 \to 11,12,20,29,   & b_3=a_3,b_3=A_3,c_3=b_3,b_4=a_4,  \\
12 \to 32,44          & c_4=A_4,d_4=b_4 \\
13 \to 9,37,42,44     & b_3=a_3,c_4=a_4,d_4=b_4,d_4=c_4 \\
14 \to 10,22,39,40    & b_3=a_3,c_3=b_3,d_4=b_4,d_4=a_4 \\
15 \to 11,12,23,24,   & b_3=A_3,b_3=a_3,c_3=B_3,c_3=b_3, \\
15 \to 37,40,43,45    & b_4=a_4,c_4=A_4,d_4=a_4,d_4=b_4 \\
16 \to 13,14,38       & c_3=a_3,b_3=a_3,c_4=a_4 \\
17 \to 27,46          & c_2=a_2,c_2=b_2 \\
18 \to 25,36,47,48,49 &b_2=a_2,c_4=a_4,c_3=b_3,c_2=a_2,c_2=A_2\\
\end{array}
$$

Arrows from level 10 to 9:
$$
\begin{array}{ll}
1 \to  2,3,4        & d_4=a_4,d_4=b_4,d_4=c_4 \\
2 \to  1,3          & c_4=a_4,c_3=b_3 \\
3 \to  1,5,6,29,34  & c_4=b_4,c_3=a_3,c_3=b_3,b_4=a_4,c_4=a_4 \\
4 \to  2,6,8,41,46  & c_4=b_4,c_3=a_3,c_3=b_3,b_4=a_4,c_4=a_4 \\
5 \to  3,6,7,40,48  & c_4=b_4,c_3=a_3,c_3=b_3,b_4=a_4,c_4=a_4 \\
6 \to  4,6,54       & c_4=b_4,c_3=b_3,c_4=a_4 \\
7 \to  9,11,17      & d_4=a_4,d_4=b_4,d_4=c_4 \\
8 \to  10,12,19     & d_4=a_4,d_4=b_4,d_4=c_4 \\
\end{array}
$$

In the table below all identifications take place in column 4 of the matrices $X_{9,i}$ and so we shall omit the subscript 4.

Arrows from level 9 to 8:
$$
\begin{array}{llll}
1 \to 2,3       & c=a,c=b      & 33 \to 17,18 & b=a,c=a \\
2 \to 3,5       & c=a,c=b      & 34 \to 18,19 & c=a,c=b \\
3 \to 3,4       & c=a,c=b      & 35 \to 8,16  & b=a,c=a \\
4 \to 3         & c=a          & 36 \to 9,18  & b=a,c=b \\
5 \to 2,16      & c=b,c=a      & 37 \to 12,16 & c=a,b=a \\
6 \to 3,15,19   & c=b,b=a,c=a  & 38 \to 10,14 & c=a,c=b \\
7 \to 4,20      & c=b,c=a      & 39 \to 10,21 & b=a,c=b \\
8 \to 5,22      & c=b,c=a      & 40 \to 15,20 & c=a,c=b \\
9 \to 6,8       & c=a,c=b      & 41 \to 19,22 & c=a,c=b \\
10 \to 7,10     & c=a,c=b      & 42 \to 18,21 & c=a,c=b \\
11 \to 8,11     & c=a,c=b      & 43 \to 15,21 & c=a,c=b \\
12 \to 9,11     & c=a,c=b      & 44 \to 14,22 & c=a,c=b \\
13 \to 6,9      & c=a,c=b      & 45 \to 18,20 & c=a,c=b \\
14 \to 7,11     & c=a,c=b      & 46 \to 15,22 & c=a,c=b \\
15 \to 8,10     & c=a,c=b      & 47 \to 14,21 & c=a,c=b \\
16 \to 6,7      & b=a,c=a      & 48 \to 19,20 & c=a,c=b \\
17 \to 6,11     & b=a,c=b      & 49 \to 10,15 & b=a,c=a \\
18 \to 6,10     & b=a,c=b      & 50 \to 10,22 & b=a,c=b \\
19 \to 7,9      & b=a,c=b      & 51 \to 10,21 & b=a,c=b \\
20 \to 7,8      & b=a,c=a      & 52 \to 10,20 & b=a,c=b \\
21 \to 7,8      & c=b,c=a      & 53 \to 15,18 & b=a,c=b \\
22 \to 13,14    & c=a,c=b      & 54 \to 15,19 & b=a,c=b \\
23 \to 8,13     & c=a,c=b      & 55 \to 14,15 & c=a,b=a \\
24 \to 12,14    & c=a,c=b      & 56 \to 10,18 & c=a,c=b \\
25 \to 8,14     & b=a,c=b      & 57 \to 10,19 & c=a,c=b \\
26 \to 7,16     & c=a,c=b      & 58 \to 12,24 & b=a,c=a \\
27 \to 15,17    & c=b,c=a      & 59 \to 13,23 & c=a,c=b \\
28 \to 7,17     & c=a,c=b      & 60 \to 17,25 & c=a,c=b \\
29 \to 15,16    & c=b,c=a      & 61 \to 16,24 & c=a,c=b \\
30 \to 7,15     & b=a,c=b      & 62 \to 1,24,26 & b=a,c=a,c=b \\
31 \to 9,12     & c=a,c=b      & 63 \to 2,26 & c=a,c=b \\
32 \to 8,13     & c=a,c=b      & 64 \to 24,26 & b=a,c=b \\
\end{array}
$$

\end{document}